# Validation and Comparison of Instrumented Mouthguards for Measuring Head Kinematics and Assessing Brain Deformation in Football Impacts

**Yuzhe Liu**[1,*,#], **August G. Domel**[1,*], **Seyed Abdolmajid Yousefsani**[1,*], **Jovana Kondic**[1,2]
**Gerald Grant**[3,4], **Michael Zeineh**[5], **David B. Camarillo**[1,3,6,]

[1] Department of Bioengineering, Stanford University, Stanford, CA, 94305, USA.
[2] Department of Electrical Engineering, Princeton University, Princeton, NJ, 08540, USA.
[3] Department of Neurosurgery, Stanford University, Stanford, CA, 94305, USA.
[4] Department of Neurology, Stanford University, Stanford, CA, 94305, USA.
[5] Department of Radiology, Stanford University, Stanford, CA, 94305, USA.
[6] Department of Mechanical Engineering, Stanford University, Stanford, CA, 94305, USA.
[*] These authors had equal contribution.
[#] Corresponding author: yuzheliu@stanford.edu

**Abstract**

Because of the rigid coupling between the upper dentition and the skull, instrumented mouthguards have been shown to be a viable way of measuring head impact kinematics for assisting in understanding the underlying biomechanics of concussions. This has led various companies and institutions to further develop instrumented mouthguards. However, their use as a research tool for understanding concussive impacts makes quantification of their accuracy critical, especially given the conflicting results from various recent studies. Here we present a study that uses a pneumatic impactor to deliver impacts characteristic to football to a Hybrid III headform, in order to validate and compare five of the most commonly used instrumented mouthguards. We found that all tested mouthguards gave accurate measurements for the peak angular acceleration, the peak angular velocity, brain injury criteria values (mean average errors < 13%, 8%, 13%, respectively), and the mouthguards with long enough sampling time windows are suitable for a convolutional neural network-based brain model to calculate the brain strain (mean average errors < 9%). Finally, we found that the accuracy of the measurement varies with the impact locations yet is not sensitive to the impact velocity for the most part.

…
*Key terms*:  concussion; mTBI; smart mouthguard; electronic mouthguard; anthropomorphic test dummy (ATD)

1. **Introduction**

In contact sports and especially in American football, concussions continue to be a major concern with nearly 4 million concussions occurring in the US alone every year [20]. Concussion is a form of mild traumatic brain injury (mTBI) resulting from rapid acceleration or deceleration of brain tissue caused by an impulsive or rotational load on the head [12]. Sports-related mTBI, which is well known as a leading cause of disability in youth [28], temporarily affects brain functionality and may result in neurodegenerative brain diseases in the long term [25]. Furthermore, growing evidence suggests that even subconcussive head impacts, if repeated, can give rise to the same neurophysiological disorders and altered MRI trajectories of brain structure [24, 26]. To better understand how sports-related head impacts affect brain health, researchers have long been studying the mechanisms underlying concussive and subconcussive impacts, and how these impacts correlate with the head kinematics and brain injury experienced during the impacts [5, 6, 18, 19, 22].

Wearable technologies in many forms have recently been equipped with sensors to measure the head kinematics during sports-related head impacts. Skin patch sensors, sensor-equipped ear plugs, the Head Impact Telemetry System (HITS – see **Table 1** for all abbreviations in this manuscript), and instrumented mouthguards are among the most common wearable technologies developed for head kinematics measurement [1, 3, 30]. Head kinematics data collected using these wearable devices has significantly shed light on the biomechanics associated with concussion [17, 25, 34, 35]. Furthermore, it was recently shown that the instrumented mouthguard has benefits over the other wearable technologies in accurately measuring head kinematics during an impact due to the rigid coupling of the upper dentition to the skull [36]. However, several studies have reported conflicting conclusions with mouthguard measurements varying from excellent accuracy [1, 3] to poor accuracy [30]. One potential explanation for these discrepancies includes the inconsistent treatment of the mandibles of the anthropomorphic test dummy (ATD) [3,16,29,30]. The mandible was fixed to clench the mouthguard with the detention in Camarillo et al. [3], while no mandible was used in Bartsch et al. [1], and finally a spring-articulated mandible was used in Siegmund et al. [30]. Another potential explanation for discrepancies among mouthguard evaluations is that the instrumented mouthguards tested in these various studies are different, and, thus, have different sensors, tightness of fit, and other design parameters leading to differences in performance. Given the importance of the instrumented mouthguards as a research tool and given these conflicting studies, there is a necessity for evaluating and comparing various common instrumented mouthguards across the same testing protocol, which we sought to carry out in this study.

Several companies and institutions have been working on further development of the instrumented mouthguards in order to supply researchers and consumers with a means of accurately measuring in-game head kinematics. However, there is still a need for a systematic performance assessment of these devices. Ensuring proper understanding of the variations among these different instrumented mouthguards and their real-time measurements is crucial for multiple reasons, including: (1) enabling meaningful insight into the underlying mechanisms of brain injury in sports-related impacts, (2) understanding the way in which these variations ultimately affect the calculation of brain deformation and brain injury criteria values, and (3) finding how these variations affect our data interpretation and how this data interpretation can guide appropriate removal (and later return) of athletes to the play in real time in the future.

In this study, we aim to address the abovementioned concerns by providing a detailed systematic validation and evaluation of five commonly available instrumented mouthguards. We use a pneumatic linear impactor and a sensor-equipped Hybrid III headform and neck to introduce impacts characteristic to football, and we then compare the kinematics obtained by the instrumented mouthguards to the reference anthropomorphic test dummy (ATD) sensors. In particular, we report how the linear acceleration, angular velocity, and angular acceleration compare with their respective references, as well as how this ultimately translates into differences in calculation of the brain strain (mechanical parameter describing the severity of deformation) and the brain injury criteria values.

2. **Materials and Methods**

The underlying assumption in mouthguard-measured head impact kinematics is the rigid coupling between the upper dentition and the skull. To assess the performance of each instrumented mouthguard, we mounted the mouthguards to a Hybrid III ATD headform by pushing the bottom of the mouthguard up and onto the upper dentition until it fit tightly onto the teeth, the same way an athlete would do so with a mouthguard during a game. We equipped the ATD with a standard football helmet (Vicis Zero1), and then conducted a series of impacts to the ATD with a pneumatic linear impactor. In addition to measuring head impact kinematics with each of the instrumented mouthguards, the ATD kinematics were also measured and analyzed for each impact. A set of high-accuracy sensors (linear accelerometers and angular velocity gyroscopes at the center of gravity of the ATD) served as the reference data (gold standard) for comparison with the instrumented mouthguard-obtained kinematics. For repeatability, three tests were performed at each of the five impact locations (facemask, front, oblique, side, and back) and four impact velocities (3.6, 5.5, 7.4, and 9.3 m/s), for each mouthguard. The mouthguard-obtained kinematic data was then processed and compared to the reference data in the following five ways: (1) the measured peak linear and angular acceleration, and angular velocity, (2) the curve correlation for the linear and angular acceleration, and angular velocity, (3) the directions of instantaneous axis, (4) the estimated brain deformation based on the impact kinematics, and (5) the predicted values of mTBI-related brain injury criteria.

### 2.1. Laboratory Setup

As shown in **Fig. 1A**, a pneumatic linear impactor (Biokinetics, Ottawa, CA) was used to introduce impacts to the helmeted ATD headform. For each impact, the velocity of the impactor head was measured immediately prior to contact with the helmet. By controlling the air pressure, we were able to achieve repeatable impact velocities. The ATD headform and the Hybrid III neck (Humanetics, Michigan, USA) were secured to a supporting table (Biokinetics, Ottawa, CA) that slides freely in the direction of the impact to mimic human body motion during an impact. We modified the height of the supporting table and rotated the ATD neck to achieve the desired impact locations.

To properly secure the chin strap of the helmet to the ATD headform, we used the modified version of Hybrid III ATD with a movable mandible (Mandible Load Sensing Headform, MLSH, Biokinetics, Ottawa, CA [29], **Fig. 1A**). The movement of the mandible is constrained by two springs and can, thus, often whip during an impact, striking the underside of the mouthguard mounted at the upper dentition [16]. This can create significant noise in the mouthguard kinematic readings. However, the occurrence of the mandible strike and the striking force have not yet been validated with human data, and since the aim of this work is to quantify and compare the accuracy of the instrumented mouthguards, we constrained the movement of the mandible to prevent this strike. This was done by placing a plate that had aluminum blocks stacked on it within the lower dentition. When the mandible whips upwards, the top of the blocks contacts the middle roof of the mouth (i.e. the middle part of the upper dentition, the blocks do not contact with the mouthguard). This stops the upward movement of the mandible before the lower dentition can strike the mouthguard. To reduce the noise upon contact, foam was attached between the blocks and roof of the mouth. Furthermore, a titanium biofidelic dentition (**Fig. 1B**) was built according to football players' representative dentition shape, which was provided by University of Pennsylvania. The simplified dentition of the MLSH was replaced with this biofidelic dentition, and all of the mouthguards were constructed to fit it. When the mouthguard is mounted to this upper dentition, the upper lip of the vinyl skin of the ATD will contact the front surface of the mouthguard. Since the vinyl skin exhibits considerably greater stiffness than that of human skin, it was observed that the vinyl skin transfers a vibration to the mouthguard causing additional noise in the mouthguard data. Therefore, part of the vinyl skin upper lip was carefully cut and modified to prevent influencing the mouthguard during the impacts.

Finally, the ATD headform kinematics were measured by a triaxial accelerometer (Dytran 3273A) at the center of gravity (CoG) as well as the three gyroscopes (DTS ARS-PRO) facing different directions. The accelerometer measured the linear acceleration at CoG, and the gyroscopes measured the angular velocity. The trigger-point of the sensors was a linear acceleration exceeding 10g in any one of the three axes. All data were acquired using the SLICE Nano & Micro software (DTS, Seal Beach, CA).

### 2.2. Mouthguards

Measuring head impact kinematics via instrumented mouthguards has shown advantage over the other sensing technologies such as, for example, skin patches, subject to relative motion between the skin and the skull [36]. To fit the teeth tightly and enable tight coupling with the skull, the instrumented mouthguards are personalized according to the user's dentition. The two common ways of achieving the mouthguard fit are the following: (1) taking an impression of user's teeth ahead of time and developing a custom mouthguard that fits their dentition, known by the name of "customized" mouthguards, and (2) heating the mouthguard by the user and biting into it to make an impression that fits their dentition, also known as the "boil-and-bite" mouthguards. In this study, we tested five mouthguards that are among the most frequently used by researchers: Stanford's customized mouthguard (denoted as MiG-C, **Fig. 2A1**), Stanford's boil-and-bite mouthguard (denoted as MiG-B, **Fig. 2B1**), Prevent Biometrics' customized mouthguard (denoted as PRE-C, **Fig. 2C1**), Prevent Biometrics' boil-and-bite mouthguard (denoted as PRE-B, **Fig. 2D1**), and, finally, Sports & Wellbeing Analytics' customized mouthguard (denoted as SWA-C, **Fig. 2E1**). The sampling time windows, time resolution, coordinate axes, and origin for each mouthguard are given in **Table 2**. For more information on all of the tested mouthguards see supporting information Section S1.

### 2.3. Testing Protocol

Head impacts in contact sports such as football can occur at different locations and various velocities. This motivated testing all of the mouthguards at five impact locations: facemask, front, oblique, side and back (**Fig. 1C-G**)), and at four velocities (3.6, 5.5, 7.4, and 9.3 m/s). The ATD headform faces parallel to the impacting direction in the facemask (**Fig. 1B**), front (**Fig. 1C**) and back (**Fig. 1D**) impacts, faces perpendicularly to the impacting direction in the side impacts (**Fig. 1E**), and faces 45° from the impacting direction in the oblique impacts (**Fig. 1F**). Regarding the impact velocities used for the testing, three of the used velocities (5.5, 7.4, and 9.3 m/s) are based on the National Football League (NFL) helmet test protocol [7], and an additional lower velocity (3.6 m/s) was added to analyze impacts of lower intensity as well. Considering that the facemask is vulnerable to failure at repeated high-speed impacts, the facemask was subjected to only the two lower impact velocities. Additionally, due to the impact velocity being controlled by a pressurized air input, the actual velocity of the impact can be slightly different from the target velocity at times. In this study, the impact velocity error of ±0.3 m/s was considered acceptable. The mean impact velocity and the standard deviation for all the tests were: 3.60±0.13 m/s, 5.50±0.08 m/s, 7.41±0.08 m/s, and 9.29±0.06 m/s.

Since the goal of this study was to compare the mouthguard-obtained kinematics to the reference ATD headform kinematics obtained in the same test impact, small variations in velocity like this are inconsequential for the study. Finally, three repetitions of each impact location and impact velocity were conducted, which resulted in 54 impacts in total (4 locations with 4 velocities and 1 location with 2 velocities, each with 3 repetitions). For consistency, we ensured that the neck was not damaged, the chinstrap was still properly fitting, and the mouthguard had not come loose before proceeding to the next impact test.

### 2.4. Data Processing and Analysis

To enable meaningful comparison between the mouthguards' and the ATD data, all mouthguard-obtained data was aligned to the ATD data and transformed to the ATD coordinate system. The details of processing ATD and mouthguard data are given in supporting information Sections (S2.1-S2.4). We used five metrics to analyze the mouthguards' performance:

(1). Relative error in the peaks of the magnitude (REPM, Eq.1): REPM assesses the accuracy of the mouthguard in measuring the peak value of the kinematics. (details in Section S2.2)

(2). Correlation coefficients of magnitude (CCM, Eq.2): CCM assesses the whole magnitude trace of the kinematics, considering that the variation of angular acceleration and the duration of the impact influence the resulting brain deformation [8, 38]. (details in Section S2.2)

(3). Instantaneous axis error (IAE, Eq.3): IAE assesses the measured direction of the impact kinematics at the peak of the magnitude, which is important considering that the brain responds differently to different impact directions [9]. (details in Section S2.2)

(4). Relative error in brain strain (REBS, Eq.4): REBS assesses the error propagating to the brain strain from the kinematics. Recently, a convolutional neural network (CNN)-based brain model for calculating the brain strain was developed and validated by Worcester head injury model (WHIM V1.0 [37], the version name of the head model comes from [39]). Unlike the finite element analysis (FEA) that takes hours to run the simulations, the CNN-based brain model can calculate the brain strain in near real-time, which has promising potential applications in the mTBI field when paired with instrumented mouthguards. To test the influence of the kinematics error on this near real-time method, in this case the CNN-based brain model, we compare the errors in the 95% maximum principal strain in the whole brain (95% MPS), 95% maximum principal strain in the corpus callosum (95% MPS at CC), and 95% fiber strain at the corpus callosum (95% FS at CC). Note that the PRE mouthguards were not assessed in REBS because applying the CNN-based brain model to these impacts leads to a large overestimation of the brain strain. (details in Section S2.3)

(5). Relative error in brain injury criteria (REBIC, Eq.4): REBIC assesses the influence of kinematic measurement errors on brain injury criteria, including Brain Angle

Metric (BAM) [17], Brain Injury Criteria (BrIC) [32], and weighted principal component score (PCS) [10] (details in Section S2.4).

$$REPM = \frac{\max(\|\overrightarrow{MG}\|) - \max(\|\overrightarrow{ATD}\|)}{\max(\|\overrightarrow{ATD}\|)} \quad \text{(Eq.1)}$$

$$CCM = \frac{\sum_{i=1}^{n}\left(\|\overrightarrow{MG}\| - \text{mean}(\|\overrightarrow{MG}\|)\right)\left(\|\overrightarrow{ATD}\| - \text{mean}(\|\overrightarrow{ATD}\|)\right)}{\sqrt{\sum_{i=1}^{n}\left(\|\overrightarrow{MG}\| - \text{mean}(\|\overrightarrow{MG}\|)\right)^2}\sqrt{\sum_{i=1}^{n}\left(\|\overrightarrow{ATD}\| - \text{mean}(\|\overrightarrow{ATD}\|)\right)^2}} \quad \text{(Eq.2)}$$

$$IAE = \arccos\left(\frac{\overrightarrow{MG}}{\|\overrightarrow{MG}\|} \cdot \frac{\overrightarrow{ATD}}{\|\overrightarrow{ATD}\|}\right) \quad \text{(Eq.3)}$$

Where $\overrightarrow{MG}$ and $\overrightarrow{ATD}$ are the vectors of the kinematics measured by the mouthguard and the ATD, respectively, at each time point. $\|\vec{x}\|$ is the magnitude of the vector $\vec{x}$. max(x) and mean(x) calculate the maximum and mean values of x over the time.

$$REBS \text{ or } REBIC = \frac{|MG - ATD|}{ATD} \quad \text{(Eq.4)}$$

Where $MG$ and $ATD$ are the brain strain or brain injury criteria values calculated using the mouthguard and ATD measured kinematics, respectively.

Linear regression is performed between the absolute values of peak kinematics given by the mouthguard and the ATD. The data is also fit to the identity line (y=x) to get a true understanding of the deviation of the mouthguards from the reference ATD. Kruskal-Wallis 1-way ANOVA is performed to compare each metric corresponding to the tested mouthguards, and p-values are calculated between every two mouthguards to see if their difference is statistically significant. Kruskal-Wallis 1-way ANOVA is, then, also used to test if a metric is a function of the impact velocity and impact location. Violin plots are used to show the distribution of data in the comparison. For each violin (the compared item), the shape of the violin is the kernel density of the data, and the box and vertical line inside the violin is simply a box and whisker plot. Moreover, the horizontal line shows the mean value of the data, and the white circle in the middle of the violin shows the median value.

### 3. Results
#### 3.1. Typical Kinematic Traces

To illustrate the head kinematics experienced during common football impacts, typical traces of angular acceleration, angular velocity, and linear acceleration at CoG in a side 9.3 m/s impact for each mouthguard are plotted in **Fig. 2.** (Traces for components were plotted in **Figs. S1-3**). In summary, peak magnitudes and whole traces matched the ATD reference for each mouthguard. Since both MiG (**Fig. 2A, B**) and SWA (**Fig. 2E**) mouthguards have relatively long sampling time windows, they demonstrated the ability to track the whole impulse, while the PRE (**Fig. 2C, D**) mouthguards missed a portion of the deceleration phase of the angular velocity. It should be noted that the angular acceleration exhibits amplified noise compared to the angular velocity in all mouthguard and ATD data due to its derivation through the differentiation of the angular velocity. Furthermore, the linear acceleration at CoG also exhibits such amplified noise since the angular acceleration is used for translating the linear acceleration from the mouthguards' sensor locations to the CoG of the headform. It is useful to note that in back impacts we observed that the bottom of the facemask may impact the neck of the ATD and generate another peak of angular acceleration after the one corresponding to the original impact. This peak was only captured by the MiG and SWA mouthguards because they have long enough sampling time windows.

### 3.2. Correlation between the Peak Magnitude of the Mouthguards and the Headform

The peak values of angular velocity, angular acceleration, and linear acceleration at the CoG given by the mouthguard and the ATD are compared in **Fig. 3** and the R-squared values and the regression equation is reported in each plot. Due to the reasons stated in Section 3.1., the data points for the angular velocity are more closely converged to *y=x*, except for the PRE mouthguards at higher angular velocities, where their gyroscopes saturated (**Fig. 3C2, D2**). The saturation of angular velocity does not, however, influence the peak angular acceleration measurements because the angular acceleration peaks prior to the angular velocity peak. It should be further noted that the outliers all correspond to a single impact location, more specifically, the front impacts' angular acceleration for MiG-B (**Fig. 3B1**), the facemask impacts' angular acceleration for Pre-B (**Fig. 3D1**), and the front and facemask impacts' linear acceleration for SWA-C (**Fig. 3E3**).

### 3.3. Assessment of Mouthguard Accuracy
#### 3.3.1. Relative Error in Peak of Magnitude (REPM)

The peaks of the magnitude of angular acceleration, angular velocity, and linear acceleration at CoG are compared in **Fig. 4**. As seen in this figure, the measured angular velocity is consistently more accurate than the angular acceleration and linear acceleration. Also, as shown in this figure, the mean REPM of angular acceleration and angular velocity for all mouthguards are smaller than 13% and 8% respectively. The mean REPM of linear acceleration at CoG for MiG mouthguards is smaller than 14%, and for PRE mouthguards it is smaller than 4%. However, the mean REPM of linear acceleration at CoG for SWA-C is 32.4%. For angular acceleration, MiG-C, PRE-C, PRE-B, and SWA-C are not significantly different, and the MiG-B, PRE-B and SWA-C have data points with slightly higher error as seen in the figure. Finally, it is interesting to note that SWA-C has the lowest error for angular velocity, MiG-C and PRE mouthguards have the lowest error for angular acceleration, and that the PRE mouthguards have the lowest error for linear acceleration at CoG.

### 3.3.2. Correlation Coefficients of Magnitude (CCM)

CCM assess the ability of the mouthguards to accurately capture the kinematic (i.e. linear acceleration, angular velocity, and angular acceleration) traces. In **Fig. 5**, we compare the CCM of the traces for each mouthguard. As shown in **Fig. 5**, the mean CCM for the MiG mouthguards are higher than 0.9, and for the PRE mouthguards the mean CCM are higher than 0.95. For SWA-C, the mean and median CCM is higher than 0.9 for angular velocity, and for linear and angular acceleration it is in the range of 0.8-0.9. It should be noted that the CCM for all mouthguards within the same developer are similar, while the CCM for the mouthguards across different developers are statistically significantly different (**Fig. 5**).

### 3.3.3. Instantaneous Axis Error (IAE)

Besides the magnitude, the ability of the mouthguards to record the accurate direction of head movement during the impact is also assessed (**Fig. 6**). We find that the mean IAEs of angular velocity, angular acceleration, and linear acceleration at CoG for all tested mouthguards are smaller than or around 10°, with MiG-C producing overall the best results. Further, the IAEs for the MiG and PRE mouthguards distribute more densely than that of SWA-C. Similarly to REPM and CCM, lower IAE can be found for angular velocity than for angular acceleration and linear acceleration.

### 3.3.4. Relative Error in Brain Strain (REBS)

To predict the amount of error in the predicted brain strain as a result of the mouthguard-obtained kinematics error, a validated CNN-based brain model [37] was applied to both the mouthguard and ATD data (**Fig. 7**). The model predicts 95% MPS in the entire brain (95% MPS), 95% MPS at the corpus callosum (95% MPS at CC), and 95% fiber strain at the corpus callosum (95% FS at CC). As shown in **Fig. 7**, the mean and median REBS for all three of these strains are lower than 10% for MiG and SWA mouthguards, and the REBS for these mouthguards are not statistically significantly different in 95% MPS and 95% FS at CC. As mentioned in the Materials and Methods Section, the PRE mouthguards data is not subjected to the brain strain analysis because its short sampling time windows makes the CNN-based brain model overestimate the brain strain.

### 3.3.5. Relative Error in Brain Injury Criteria (REBIC)

Ultimately, the head kinematics are used to calculate the brain deformation. We use three representative mTBI risk criteria (BAM, BrIC, and PCS) to determine the extent of the error in the mouthguard kinematics that translates to error in the mTBI risk criteria.

The predicted values are calculated based on the mouthguard and the ATD data and compared in **Fig. 8**. The mean and median values of relative error in BAM, BrIC, and PCS for all mouthguards are below 13%. Furthermore, the lowest error is found in BrIC because the inputs for BrIC are the angular velocities, which are the most accurate of all kinematic measurements as seen in **Fig. 4B**. The inputs for BAM are the whole traces of angular acceleration, and the inputs for PCS are the traces of linear acceleration and the peak of angular acceleration. Thus, higher errors are seen in the results for the BAM and PCS criteria. Finally, it is interesting to note that MiG-C and PRE-C had statistically identical and the best overall performance for REBIC.

### 3.4. Assessment at Different Impact Locations

In our testing protocol, the ATD was impacted at five different locations, and the errors in the various metrics for different impact locations are compared in **Figs. S5-S9**. Note and PRE mouthguards are not compared in REBS because the CNN-based model will overestimate its brain strain. To further study how the accuracies of the mouthguard measurements rely on all of the various impact locations, the statistical significances of the differences are listed in **Table 3**. Smaller p-values indicate that the difference of metrics at impact locations is more significant, and we define the metric is highly related to the impact locations for $p<0.001$. For the MiG mouthguards, most of the metrics highly rely on the impact locations, while the REBIC depends less on impact location for MiG-C than it does for MiG-B. For the PRE mouthguards, the metrics for PRE-C are not highly related to impact location except for CCM of angular acceleration, while most of the metrics for PRE-B are highly related to impact location. In regard to SWA-C, the various metric errors are not as highly related to the impact location as is observed in the MiG-B and PRE-B. It can thus be concluded that the accuracy of the boil-and-bite mouthguards generally tends to be more dependent on the impact location.

### 3.5. Assessment at Different Impact Velocities

At each impact location, the ATD was impacted at four impact velocities except for the facemask impacts, where only the two lower velocities were used. The performance metrics of mouthguards are compared in **Figs. S10-S14** and the p-values are listed in **Table 4** to show the dependence of a given metric on impact velocity. Smaller p-values indicate that the metric relies more on the impact velocity and we define the metric highly related to the impact velocity for $p<0.001$. Similar to Section 4.1, the PRE mouthguards are not compared in REBS due to their time window being shorter than required for the CNN-based brain model. Additionally, considering that the facemask is susceptible to failures at high impact velocities, only the two lower impact velocities were used for the facemask impacts, as discussed in the Materials and Methods Section. Thus, to avoid bias, the impacts at the facemask were not incorporated in this analysis of impact velocity.

As shown in **Table 4**, the CCM of the linear acceleration at CoG for MiG-C relies highly on the impact velocity as evidenced by the lower CCM values at the lower impact velocities (**Fig. S11)**. For the PRE mouthguards, the REPM of angular velocity (**Fig. S10**) and the CCM of angular velocity and linear acceleration at CoG (**Fig. S11**) rely highly on the impact velocity as evidenced by the saturation of the angular velocity of these mouthguards at higher impact velocities (**Fig. 2 C2, D2**). On the other hand, as seen in **Table 4**, the metrics for MiG and SWA mouthguards are not highly related to the impact speeds. Furthermore, it should be noted that most of the REBS and REBIC measures do not rely on the impact velocity. This indicates that conducting further analysis on the measured kinematics is not highly influenced by the impact velocity.

## 4. Discussion

In this study, we assessed the accuracy of five instrumented mouthguards from three mouthguard developers with metrics considering the peaks, the traces, and the directions of the kinematics. Despite the differences among the five tested mouthguards, the accuracy of all measured kinematics (R-squared values in **Fig. 3**) is in close agreement with previous studies in which the mandible strikes were also prevented [1, 3, 27]. However, the accuracy of the mouthguard-obtained kinematics is higher than what was observed in prior work in which the spring-driven mandible strike was allowed [30]. The difference between this current study and this previous study [30] is that the direct impact from the mandible to the mouthguard led to mouthguard loosening or large local deformation of its mouthguard material. As a result, relatively high error was observed in this previous study [30]. In future work, the actual occurrence and the level of mandible strike loading should be quantified and carefully investigated to further understand its implications in human studies.

In general, the accuracy of the various mouthguards varies with the types of kinematics considered. As shown in **Fig. 4**, the REPM of angular velocity is more accurate than angular acceleration and linear acceleration at CoG because the derivative amplifies noise in the calculation of angular acceleration. Since the linear acceleration requires the angular acceleration in calculating its value at CoG, the angular acceleration error further propagates to the linear acceleration at CoG. Regardless, the accuracy of angular velocity may be more important to the study of mTBI because both animal [4, 11] and human [23, 25] studies have demonstrated that the angular velocity is a promising kinematic parameter to predict mTBI, and FEA simulations suggest that the peak of the angular velocity correlates well with MPS [17].

Besides the peak value, the time variations of the kinematics, especially the angular velocity, is further important for evaluating brain deformation [8]. We found that the time variation of mouthguard-measured angular velocity is accurate across the various mouthguards tested (CCM > 0.92, **Fig. 5**) and can provide a reliable input for analysis of brain MPS (**Fig. 7**) for head models that take angular velocity traces as an input [40].

In addition to the time history of the kinematics, we also considered the direction of the kinematics. Although prior work showed that the direction of head kinematics is not influential to the brain MPS [15], it might affect the brain deformation at a specific region, and especially might affect the results of a detailed brain model [21] used to investigate pathologies observed by magnetic resonance imaging and histopathology. We found that the mean IAE of angular velocity of the various mouthguards ranges from 3° to 10° (**Fig. 6**), and we suggest that its influence on the deformation of specific brain regions should be further investigated before correlating the pathology to the regions of high MPS.

Since the measured kinematics are typically used to calculate the brain strain and to assess potential risk of injury, we further calculated brain strain (**Fig. 7**) and brain injury criteria (**Fig. 8**) based on the measured kinematics. We found that the REBS and REBIC are lower than or close to the REPM for the tested mouthguards. This indicates that the CNN-based brain model and the calculation of the brain injury criteria would not further amplify the error in the measured head kinematics (and would even potentially suppress it). The obtained accuracies of all mouthguards that were tested in this study are, thus, evaluated as adequate for further analyses of the kinematic measurements via the CNN-based brain model. This is an important guideline for studies that would benefit from a method of calculating MPS that is substantially quicker than traditional finite element brain models.

In addition to understanding the errors associated with the measured kinematics and the further brain analyses, it is important to understand how the measurement error in the kinematics varies with impact location and velocity. Interpreting measured kinematics with respect to the impact location has already been indicated by previous studies [3, 30] that observed accuracy varying with impact location. As seen in **Figs. S10-S14**, except for the PRE-C mouthguard, the kinematic measurements of facemask impacts are generally less accurate than in other impact locations. This reduced accuracy of facemask impacts has already been observed in [1], and is due to the propagation of the loading that causes noise greater than in impacts introduced directly to the shell of the helmet. Besides different pathways for loading propagation, the variation of accuracy at different locations could be explained by the impact location-dependent inertial force on the mouthguard, as well as the resistance to mouthguard loosening is also location-dependent. The key to an instrumented mouthguard's accuracy in measuring head kinematics is the rigid coupling between the upper dentition and the skull. Therefore, besides the errors inherent to the sensors, the accuracy of mouthguard-obtained kinematic measurements is heavily influenced by the relative motion between the mouthguard and the upper dentition. Frictional force between the teeth and the inner surface of the mouthguard is sometimes overcome by the inertial force experienced during an impact, which further leads to noisy measurements as a result of the relative motion (loosening) between the mouthguard and the dentition. The magnitude and direction of inertial force vary because the head-neck system responds differently to impacts in the sagittal and coronal planes [14]. In addition to this, it is observed in **Fig. S4** that the angular acceleration contribution to the linear acceleration at CoG depends on the impact location. As a result, the propagation of error from angular acceleration to the linear acceleration at CoG also depends on the impact location. Furthermore, comparing **Figs. S4A, B** and **Figs. S4C, D,** it should be noted that the contribution of angular acceleration to linear acceleration at CoG for the same impact location varies with the location of the gyroscope among the mouthguards. Finally, comparing the customized and the boil-and-bite mouthguards in relation to impact locations, customized mouthguards consistently appear to have fewer metrics with significant dependence on the impact location compared to the boil-and-bite mouthguards. This could likely be due to a more sophisticated fit to the dentition compared to the boil-and-bite method.

With regards to the effect of impact velocity on mouthguards' accuracy, the main challenge here is the lack of a feasible way of quantifying impact velocities in on-field games, thus making it difficult to update the interpretation of the head kinematic measurements accordingly. Therefore, it is critical for the instrumented mouthguards to be able to accurately measure impacts across a wide range of impact severities. As the impact velocity increases, the impacted head moves faster and the inertial forces pulling the mouthguard off the dentition can be higher. As a result, relative motion between the mouthguard and dentition is more likely to occur at higher impact velocities. In addition to the mechanical factors associated with higher velocity impacts, the gyroscope itself can degrade mouthguards' accuracy by saturating at high angular velocities that are reached as a result of high-velocity impacts. However, even though some of the metrics from Section 3.3 depend on impact velocity, the majority of the metrics do not exhibit a significant dependency on impact velocity (**Table 4**).

Besides understanding and analyzing the instrumented mouthguards as a whole, it is important to also understand and interpret the results of each of the types of various mouthguards individually. Regarding the MiG mouthguards specifically, we found that the measurement accuracy did not highly rely on the impact velocity due to a measurement range that is large enough to avoid gyroscope saturation (see **Table 4**). Moreover, we found that the side impacts, in accordance with [30], yield the most accurate measurements, whereas the facemask impacts as well as the front impacts for MiG-B lead to relatively larger errors. We further observed an underestimation of the angular acceleration and linear acceleration and a slight overestimation of the angular velocity (**Fig. 4**). Compared with the other mouthguards, the MiG mouthguards exhibit a higher mean REPM. Specifically, in a small group of impacts measured by MiG-B, REPM is higher than 20% (**Fig. 4A**). However, the performance of the MiG mouthguards in CCM and IAE is similar to the PRE and SWA mouthguards, which results in a relatively similar performance in the REBS and REBIC. The reason for the relatively higher REPM of MiG mouthguards can be explained by the greater measurement ranges in the MiG's accelerometer (400 g) and gyroscope (70 rad/s) as shown in **Table 1**. The full measurement range of the MiG mouthguards was not utilized in this testing protocol, and on-field football data [12] suggests that the gyroscope measurement range of 35 rad/s will be adequate for the majority of impacts. However, the angular velocities of a few impacts are beyond 35 rad/s [12]. Therefore, the MiG mouthguards (or other mouthguards with wide measurement ranges) are important in capturing these most dangerous impacts. Since the linear acceleration at CoG is not directly measured by the accelerometer and tends to not play a significant role in calculating brain deformation in mTBI head impacts [17], we do not discuss the accelerometer measurement range.

Regarding the PRE mouthguards specifically, we found that PRE mouthguards are very good at tracking the linear acceleration at CoG, including the slight fluctuations after the peak value that were not measured accurately by the other mouthguards (comparing **Fig. 2 A4, B4, C4, D4, E4)**. We found the REPM to be significantly lower and CCM significantly higher compared to the MiG and SWA mouthguards (**Fig. 4**), while the IAE are comparable with the MiG mouthguards and lower than the SWA mouthguard (**Fig. 5**). We further found that PRE mouthguards provide accurate measurements of the side impacts (as observed in the MiG mouthguards and in previous work [30]), and only yield relatively larger errors in facemask impacts. The errors in facemask impacts for the PRE-C mouthguard, however, were still relatively small, which may be explained by the tighter fit of the customized mouthguard to the dentition that was noticed during testing. We also observed the PRE mouthguards' accuracy varying with the different impact velocities (see **Table 4**), particularly as a result of saturation of the PRE-C mouthguard's gyroscope (**Fig. 2 C3, Fig. S2C**). While this saturation was only observed for angular velocities exceeding 35 rad/s in a single direction (**Table 1**) which is generally rare in football [12], this should still be taken into account when considering the most severe cases or utilizing the mouthguards in sports that introduce potentially higher angular velocities. No saturation was found for the PRE-B mouthguard (**Fig. 2 D3, Fig. S2D**), as a result of the impact direction being different from the measuring directions of the gyroscope, and amplitude of each component, thus, being lower than the saturation limit even when its magnitude is not. It should be noted, however, that the PRE-B mouthguard may still saturate when the impact direction changes in a way that aligns more closely with its gyroscope's measuring directions. Lastly, as a result of short sampling time windows (**Table 1**), we found that the PRE mouthguards did not fully capture the deceleration phase of the angular velocity (as seen in **Fig. 2 C3 and D3**), which may explain the overestimation of the brain strain given by feeding the PRE data into the CNN-based brain model. Considering that the kinematics of lab head impacts may differ from on-field impacts, we are investigating the adequacy of sampling time windows and triggering conditions for on-field football impacts in another study.

Regarding the SWA mouthguard, it is important to note that it was originally designed and tuned for rugby, in which the head kinematics may be differ from that of football. The SWA mouthguard provides accurate measurements of the angular acceleration and angular velocity but overestimates the linear acceleration at CoG (**Figs. 2 E5, 3 E3**). However, in spite of the larger errors associated with the linear acceleration at CoG, the REBS (time windows of 100 ms is enough for the CNN-based brain model) and REBIC of the SWA mouthguard are similar compared to the other mouthguards (**Figs. 7, 8**) since linear acceleration does not influence the brain deformation [17]. Similarly to the MiG and PRE mouthguards, the facemask impacts yielded measurements that were most prone to errors compared to the other impact locations. It is also interesting to note that, in spite of an equivalent measuring range of the gyroscopes in the SWA and PRE mouthguards (**Table 1**), saturation was not observed in the SWA mouthguards (**Fig. 2 E3, Fig. S2E**) because of the misalignment between the mouthguard gyroscope's measuring directions and the impact directions (similar to reason why PRE-B did not saturate). Furthermore, it should be noted that the MiG and PRE mouthguards have sensors at the incisor, while the SWA mouthguard has a sensor at the molar, which can be more susceptible to the mandible strike as was suggested in prior work [16]. This different positioning of the sensors also explains why the distribution of error in SWA mouthguard measurements differs from the way in which errors in MiG and PRE mouthguards are distributed.

In summary, we found that all tested mouthguards demonstrate sufficient accuracy in measuring the head kinematics of football impacts, and mouthguards with sufficiently long sampling time windows are also adequate for further brain strain analyses using the CNN-based brain model. The MiG mouthguards are particularly suitable for studies focusing on severe head impacts because of their wide measurement ranges. Additionally, the relatively long sampling time windows of the MiG mouthguards enable them to capture any potential kinematic peaks that occur considerably later than the trigger. On the other hand, the PRE mouthguards exhibit high measurement accuracy, especially in linear acceleration, and are, thus, suitable for studies that necessitate particularly high kinematic measurement accuracy. We advise, however, that researchers considering the PRE mouthguards make a prior estimate of the sampling time windows of the data required for their analyses. The SWA mouthguard was found to fail in providing reliable measurements of the linear acceleration at CoG, but has provided acceptable measurements of the rotational kinematics, thus, making it suitable for further brain deformation analysis with the CNN-based model. Lastly, comparing the two kinds of mouthguards within the same developer, we noticed that the customized mouthguards generally tend to be more accurate than the boil-and-bite mouthguards as a result of their tight fitting to the dentition. Most of the performance metrics, however, do not differ significantly between the two kinds of mouthguards, indicating that the boil-and-bite mouthguards can still be useful for studies in which it is crucial to have a quicker way of fitting the mouthguard to the players' dentitions.

Finally, it is important to note the potential limitations. One limitation of this study is that the mandible strike is not considered. In two studies using the ATD with an unconstrained movable mandible [16, 29], the mouthguard-obtained data was consistently higher in magnitude than the ATD data. This can be explained by the impact introduced to the mouthguard from the mandible strike. In the present study, the influence of the mandible strike was not considered, and we instead fixed the articulated spring-constrained mandible to prevent direct strikes to the mouthguard and allow for a cleaner assessment and comparison of the various mouthguards' accuracy. Therefore, quantifying the occurrence of mandible strikes and their resulting loading is needed in future work, especially in human studies. Besides, the noise caused by the mandible strike is in part a result of the stress wave propagation through the mouthguard during the strike. The resulting reading may have different characteristics from the rigid movement of the skull, and we think it may be important to develop an algorithm to detect a mandible strike in on-field play in future work, in order to further understand its importance in human studies. Additionally, the mouthguard developers should carefully select their material to suppress the stress propagation and potentially incorporate more energy absorption into the mouthguard design. Another limitation of this study includes not considering football ground-head impacts (for example via a drop test), which may yield different responses due to the higher stiffness of the ground compared to the impactor head. Moreover, the REBS was assessed based on lab impacts, and the actual REBS in on-field data may differ because of the different kinematic characteristics of lab impacts versus on-field impacts. Finally, one additional limitation is that only three of the common instrumented mouthguard developers (and five types of mouthguards in total) were tested, while there are other companies and institutions that have developed instrumented mouthguards and customized mouthpieces [13, 27, 31, 33]. In the future, besides the accuracy, it is important to test the mechanical safety associated with the embedded electronics of the different mouthguards, as done in previous work [2].

## Acknowledgements

This research was supported by the Pac-12 Conference's Student-Athlete Health and Well-Being Initiative, the National Institutes of Health (R24NS098518), Taube Stanford Children's Concussion Initiative, and the Office of Naval Research Young Investigator Program (N00014-16-1-2949). The content of this article is solely the responsibility of the authors and does not necessarily represent the official views of funding agencies. It should be noted that the MiG mouthguard and intellectual property associated with it is owned by Stanford University. We would like to thank Dr. Michael Fanton, Dr. Hossein Vahid Alizadeh, and Dr. Sarah McGough for the insightful discussions on the development of the experimental protocol and data analysis. We also want to thank Prevent Biometrics and SWA companies for providing the mouthguards and answering any necessary questions

20. Langlois J. A., W. Rutland-Brown and M. M. Wald. The epidemiology and impact of traumatic brain injury: a brief overview. *The Journal of head trauma rehabilitation* 21: 375-378, 2006.
21. Li X., Z. Zhou and S. Kleiven. An anatomically accurate and personalizable head injury model: Significance of brain and white matter tract morphological variability on strain. *bioRxiv* 2020.2005.2020.105635, 2020.
22. Lu Y.-C., N. P. Daphalapurkar, A. Knutsen, J. Glaister, D. Pham, J. Butman, J. L. Prince, P. Bayly and K. Ramesh. A 3D computational head model under dynamic head rotation and head extension validated using live human brain data, including the falx and the tentorium. *Annals of Biomedical Engineering* 47: 1923-1940, 2019.
23. Margulies S. S. and L. E. Thibault. A proposed tolerance criterion for diffuse axonal injury in man. *Journal of Biomechanics* 25: 917-923, 1992.
24. Miller L. E., E. K. Pinkerton, K. C. Fabian, L. C. Wu, M. A. Espeland, L. C. Lamond, C. M. Miles, D. B. Camarillo, J. D. Stitzel and J. E. Urban. Characterizing head impact exposure in youth female soccer with a custom-instrumented mouthpiece. *Research in Sports Medicine* 28: 55-71, 2020.
25. O'Keeffe E., E. Kelly, Y. Liu, C. Giordano, E. Wallace, M. Hynes, S. Tiernan, A. Meagher, C. Greene and S. Hughes. Dynamic Blood–Brain Barrier Regulation in Mild Traumatic Brain Injury. *Journal of Neurotrauma* 37: 347-356, 2020.
26. Parivash S. N., M. Goubran, B. D. Mills, P. Rezaii, C. Thaler, D. Wolman, W. Bian, L. A. Mitchell, B. Boldt and D. Douglas. Longitudinal changes in hippocampal subfield volume associated with collegiate football. *Journal of Neurotrauma* 36: 2762-2773, 2019.
27. Rich A. M., T. M. Filben, L. E. Miller, B. T. Tomblin, A. R. Van Gorkom, M. A. Hurst, R. T. Barnard, D. S. Kohn, J. E. Urban and J. D. Stitzel. Development, validation and pilot field deployment of a custom mouthpiece for head impact measurement. *Annals of Biomedical Engineering* 47: 2109-2121, 2019.
28. Seifert T. D. Sports Concussion and Associated Post-Traumatic Headache. *Headache: The Journal of Head and Face Pain* 53: 726-736, 2013.
29. Siegmund G. P., K. M. Guskiewicz, S. W. Marshall, A. L. DeMarco and S. J. Bonin. A headform for testing helmet and mouthguard sensors that measure head impact severity in football players. *Annals of Biomedical Engineering* 42: 1834-1845, 2014.
30. Siegmund G. P., K. M. Guskiewicz, S. W. Marshall, A. L. DeMarco and S. J. Bonin. Laboratory validation of two wearable sensor systems for measuring head impact severity in football players. *Annals of Biomedical Engineering* 44: 1257-1274, 2016.
31. SISU. SISU Sense. https://www.sisuguard.com/sisu-sense-electronic-mouth-guard/.
32. Takhounts E. G., M. J. Craig, K. Moorhouse, J. McFadden and V. Hasija. Development of brain injury criteria (BrIC). SAE Technical Paper, 2013.
33. Force Impact Technologies. FITGuard. https://fitguard.me/.
34. Tiernan S., A. Meagher, D. O'Sullivan, E. O'Keeffe, E. Kelly, E. Wallace, C. P. Doherty, M. Campbell, Y. Liu and A. G. Domel. Concussion and the severity of head impacts in mixed martial arts. *Proceedings of the Institution of Mechanical Engineers, Part H: Journal of Engineering in Medicine* 0954411920947850, 2020.
35. Wu L. C., C. Kuo, J. Loza, M. Kurt, K. Laksari, L. Z. Yanez, D. Senif, S. C. Anderson, L. E. Miller and J. E. Urban. Detection of American football head impacts using biomechanical features and support vector machine classification. *Scientific Reports* 8: 1-14, 2017.
36. Wu L. C., V. Nangia, K. Bui, B. Hammoor, M. Kurt, F. Hernandez, C. Kuo and D. B. Camarillo. In vivo evaluation of wearable head impact sensors. *Annals of Biomedical Engineering* 44: 1234-1245, 2016.
37. Wu S., W. Zhao, K. Ghazi and S. Ji. Convolutional neural network for efficient estimation of regional brain strains. *Scientific Reports* 9: 1-11, 2019.
35

# Figures and Tables for Manuscript

## Figures:

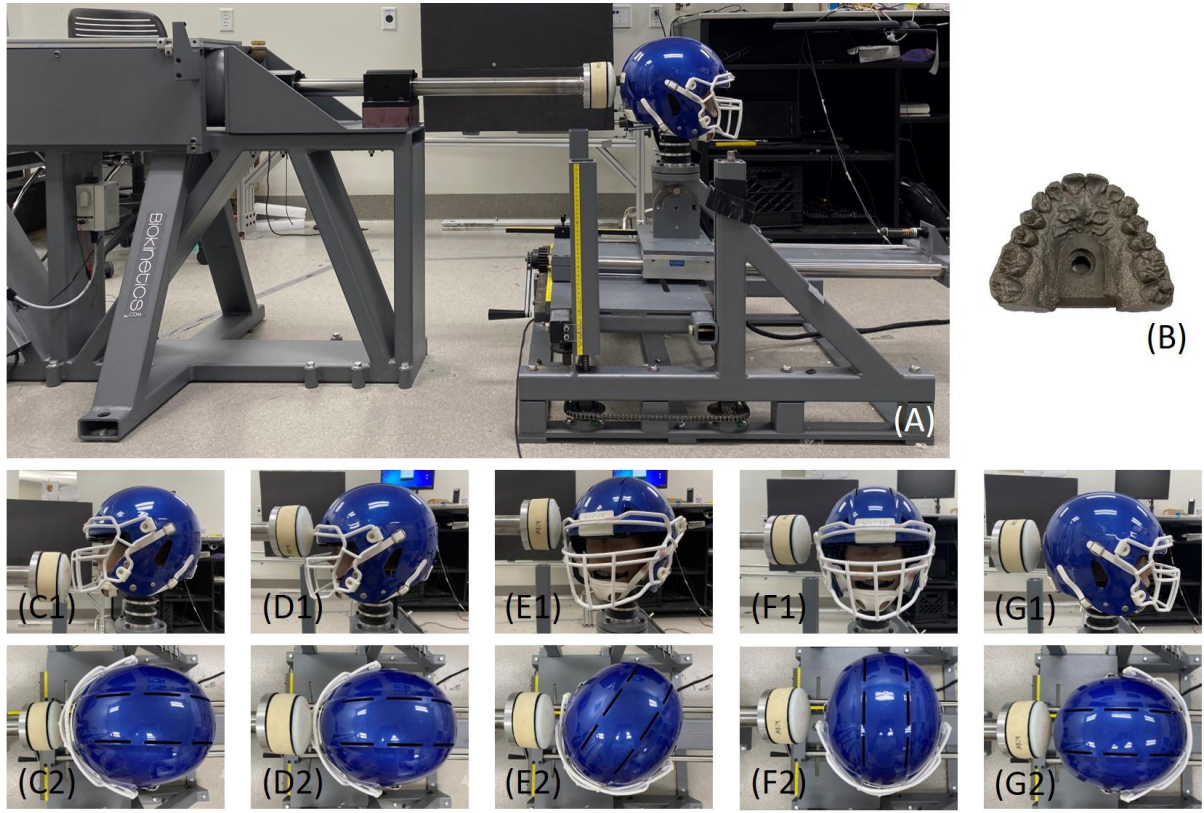

**Figure 1.** (A) Pneumatic linear impactor, supporting table, and helmeted anthropomorphic test dummy (ATD); (B) titanium biofidelic dentition; (C) Facemask impact setup: (C1) Side view, (C2) Top view (Similar to D-F) ; (D) Front impact setup; (E) Oblique impact setup; (F) Side impact setup; (G) Back impact setup.

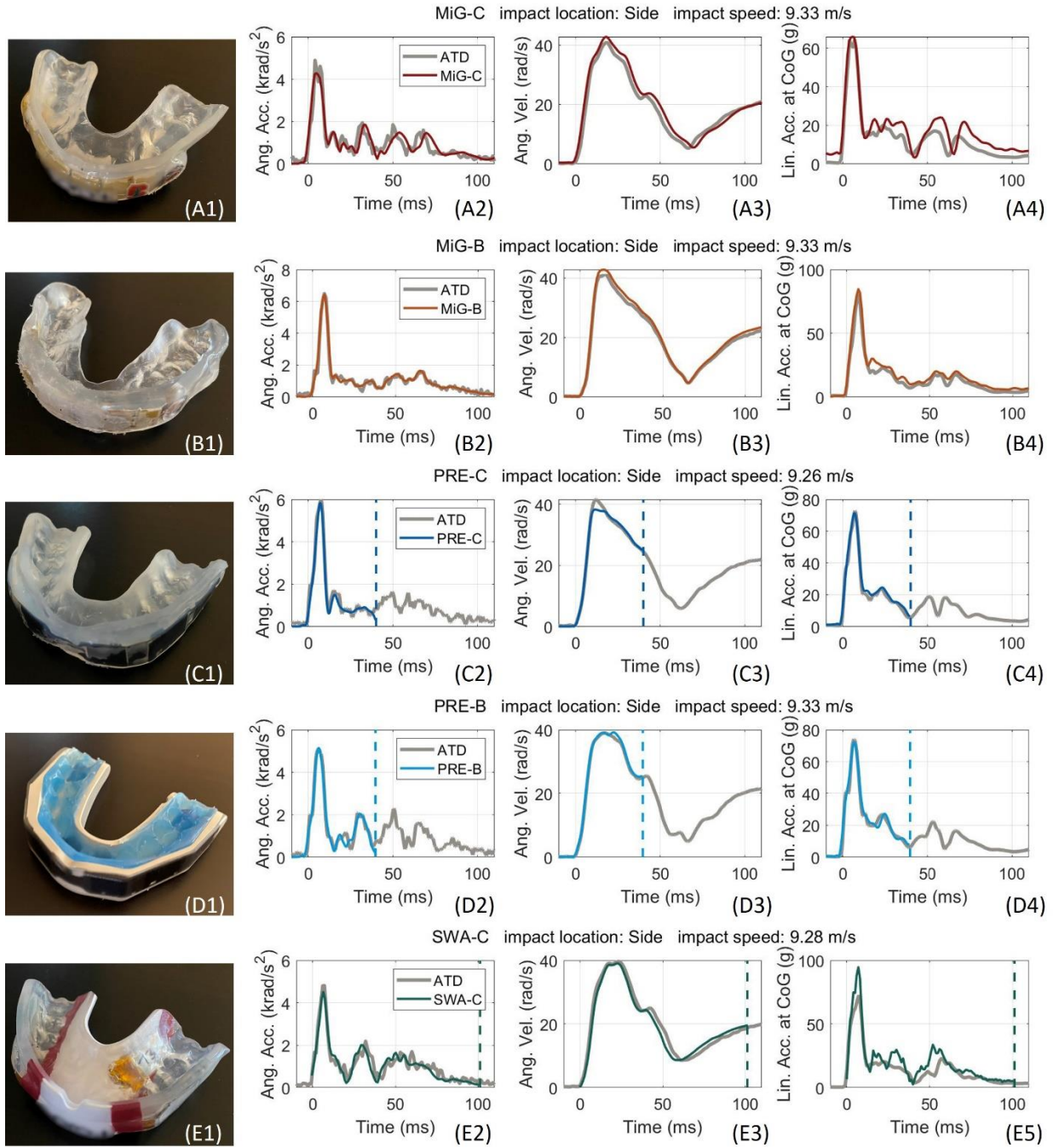

**Figure 2.** Magnitude of the angular acceleration, angular velocity, and linear acceleration at CoG for ~9.3 m/s side impact. (A) MiG-C, (B) MiG-B, (C) PRE-C, (D) PRE-B, (E) SWA-C; (1) Instrumented mouthguard configuration, (2) Angular acceleration, (3) Angular velocity, (4) Linear acceleration at CoG. Dash lines in C, D and E denote the end of the measurement time windows. The traces for components are in supporting information Section S2 (**Figs. S1-S3**).

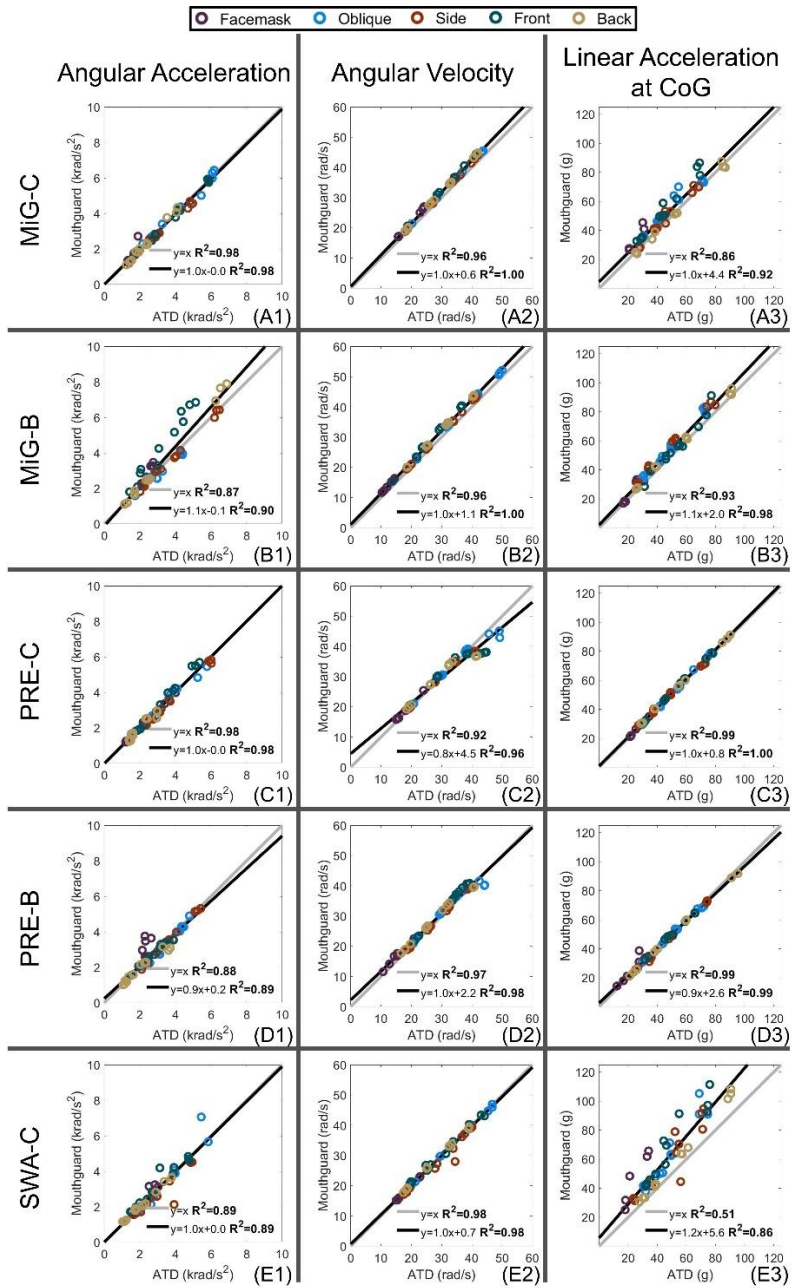

**Figure 3.** Comparison of the peak values of angular acceleration, angular velocity, and linear acceleration between the mouthguard and the ATD. A linear regression is performed on the ATD and the mouthguard data (black line). The obtained function and corresponding R-squared value are reported in the legend. Additionally, the data is fit to the identity function (*y*=*x*; gray line) in order to allow for a direct comparison of the deviation of the mouthguard data from the reference data. The R-squared of this fitting is reported in the legend as well. Each row corresponds to a mouthguard: (A) MiG-C, (B) MiG-B, (C) PRE-C, (D) PRE-B, (E) SWA-C; and each column corresponds to the components of the kinematics: (1) Angular acceleration, (2) Angular velocity, (3) Linear acceleration at CoG.
39

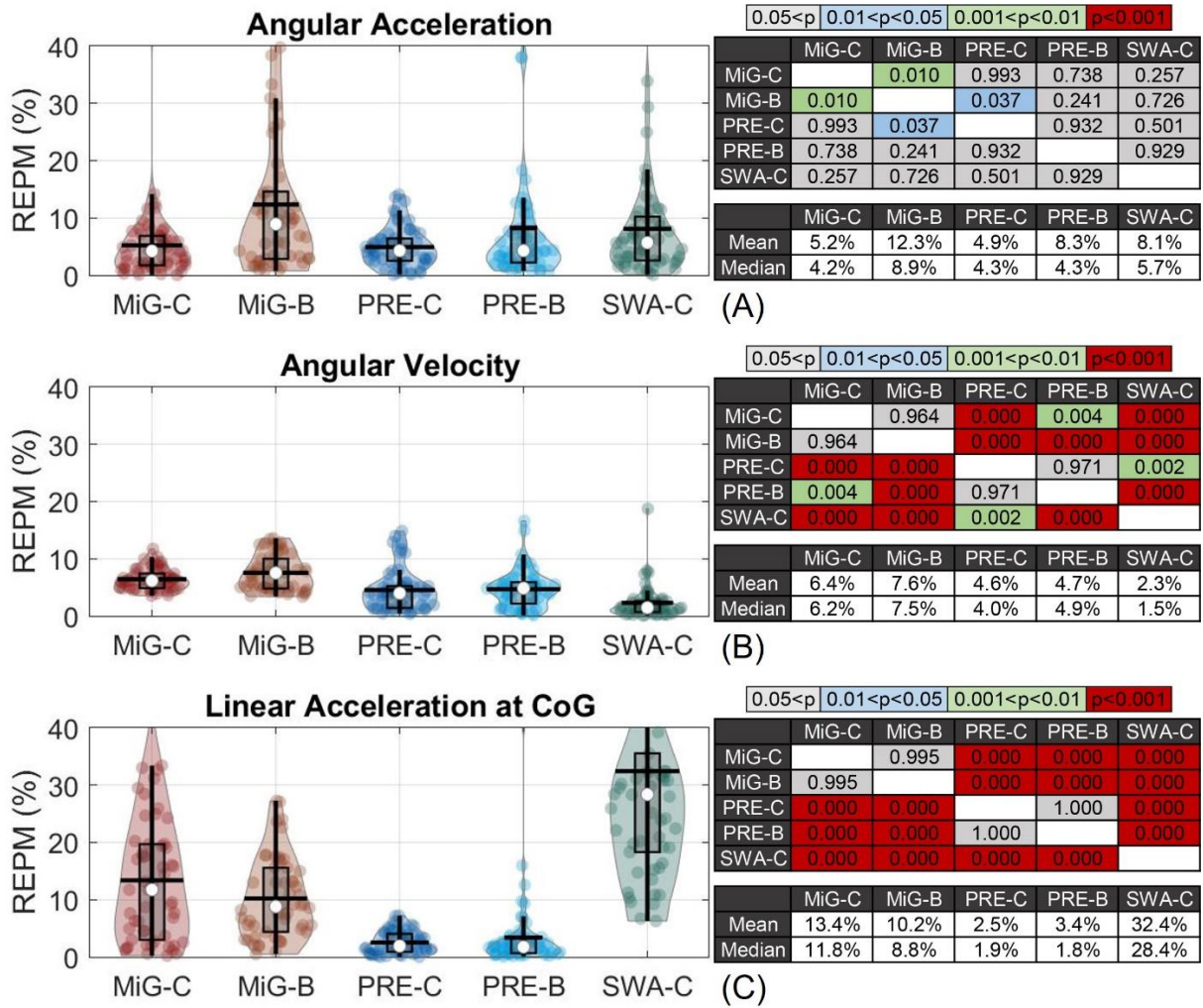

**Figure 4.** Relative error in the peak of the magnitude (REPM) of (A) angular acceleration, (B) angular velocity, and (C) linear acceleration at CoG for each instrumented mouthguard. In each row of the figure, the table on the top-right displays the p-values of the Kruskal-Wallis 1-way ANOVA test between the REPM of every two mouthguards, and the table on the bottom-right lists the mean and median REPM of each mouthguard.

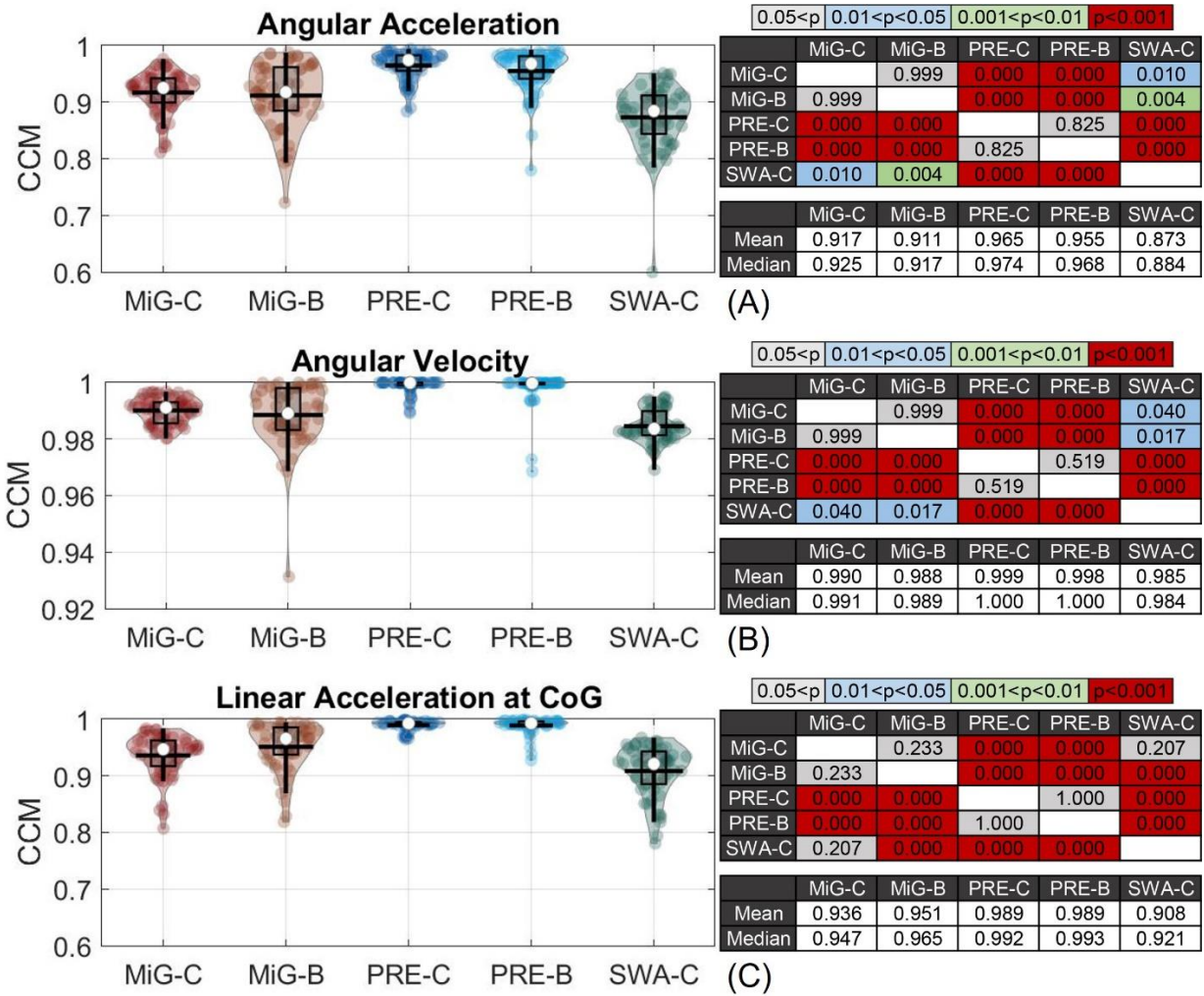

**Figure 5.** Correlation coefficient of the magnitude (CCM) of (A) angular acceleration, (B) angular velocity, and (C) linear acceleration at CoG for each mouthguard. In each row of the figure, the table on the top-right displays the p-values of the Kruskal-Wallis 1-way ANOVA test between the CCM of every two mouthguards, and the table on the bottom-right lists the mean and the median CCM of each mouthguard.

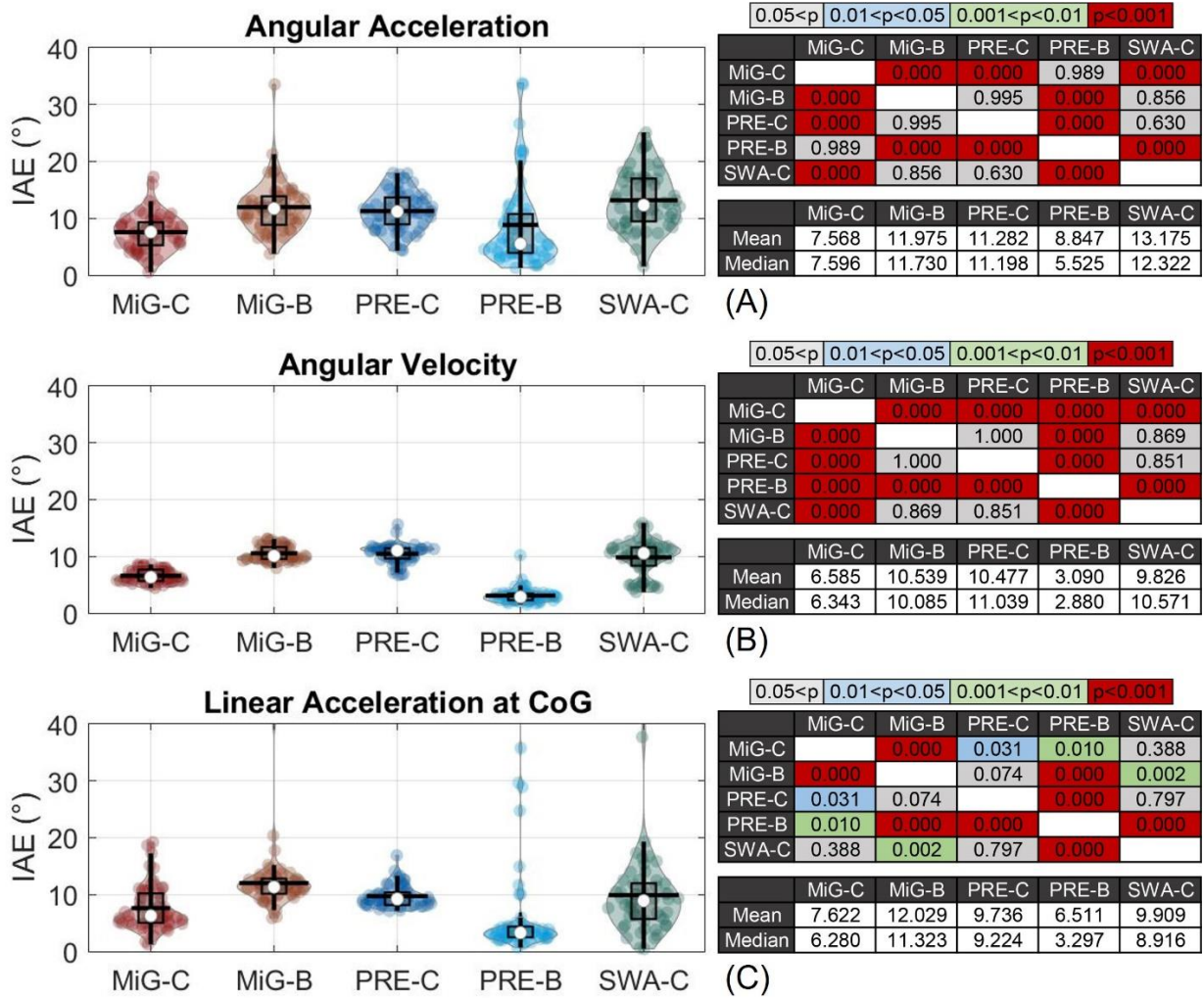

**Figure 6.** Instantaneous axis error (IAE) of (A) angular acceleration, (B) angular velocity, and (C) linear acceleration at CoG at the time corresponding to the peak of the magnitude, given by the ATD. In each row of the figure, the table on the top-right displays the p-values of the Kruskal-Wallis 1-way ANOVA test between the IAE of every two mouthguards, and the table on the bottom-right lists the mean and the median IAE of each mouthguard.

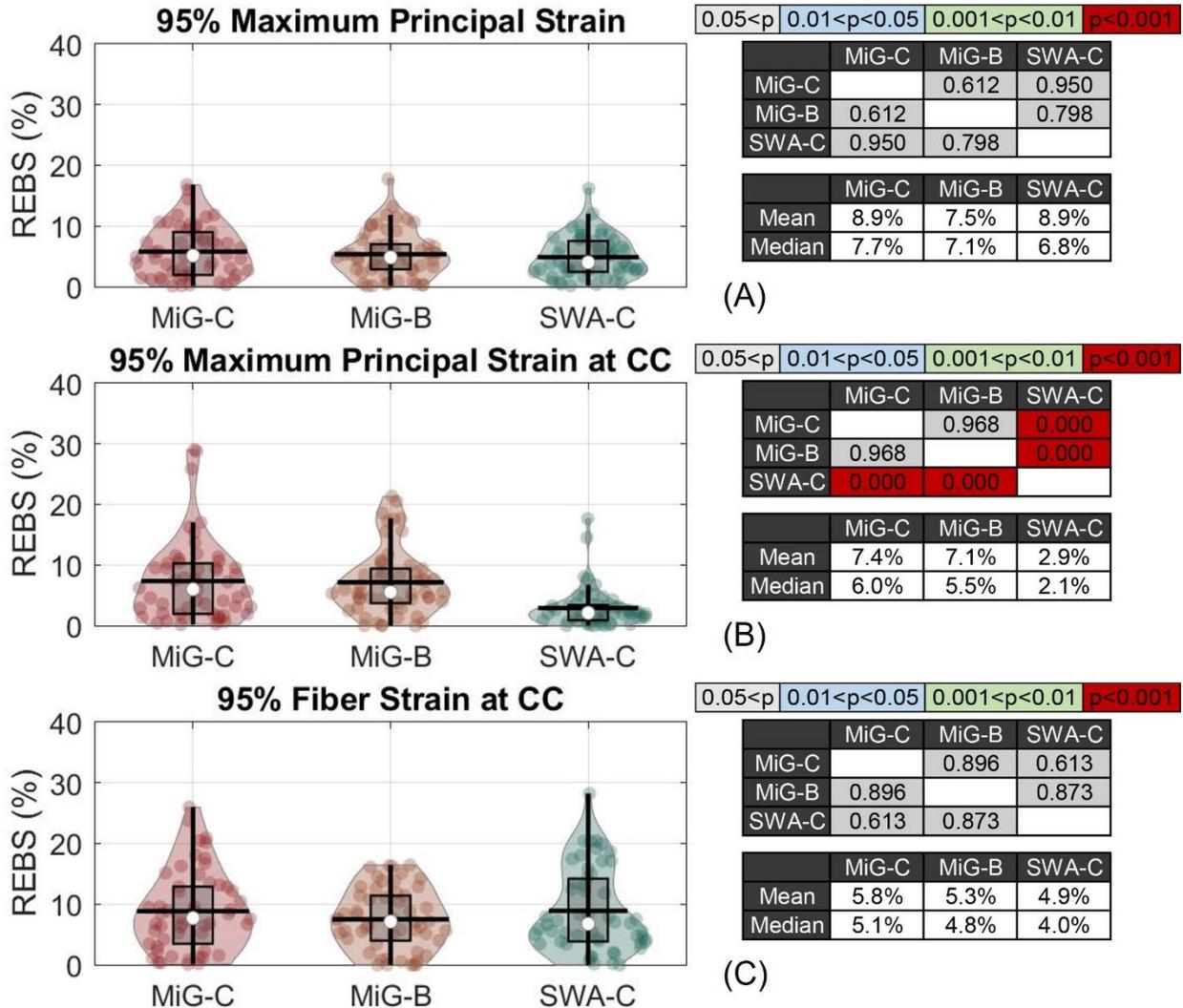

**Figure 7.** Relative error in brain strain (REBS) for each mouthguard. In each row of the figure, the table on the top-right displays the p-values of the Kruskal-Wallis 1-way ANOVA test between the REBS of every two mouthguards, and the table on the bottom-right lists the mean and the median REBS of each mouthguard. (A) 95% Maximum principal strain (95% MPS). (B) 95% Maximum principal strain at the corpus callosum (95% MPS at CC). (C) 95% Fiber strain at the corpus callosum (95% FS at CC). The PRE mouthguards are not compared in REBS because their time windows are shorter than required.

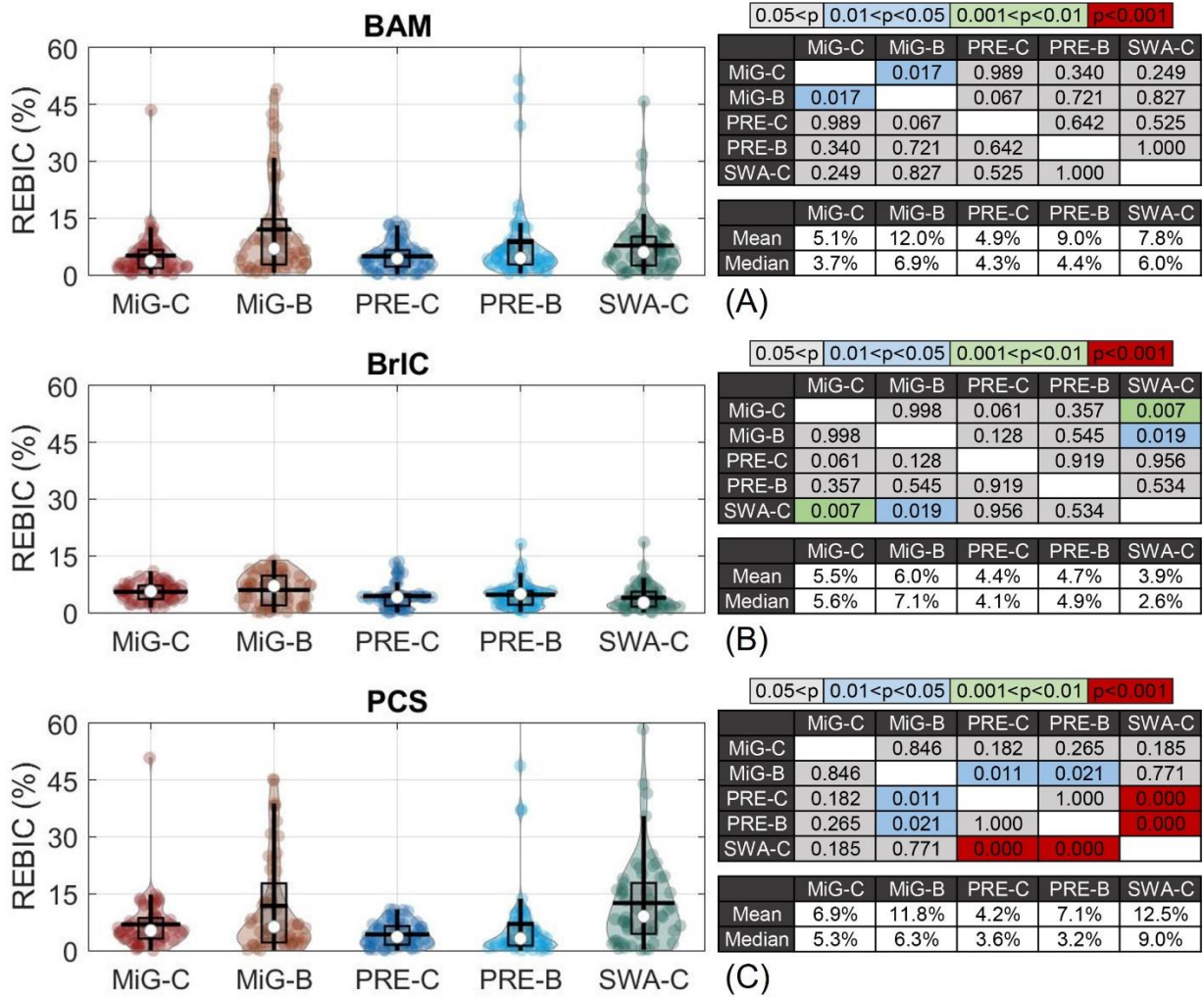

**Figure 8.** Relative error in brain injury criterion (REBIC) for each mouthguard. In each row of the figure, the table on the top-right displays the p-values of the Kruskal-Wallis 1-way ANOVA test between the REBIC of every two mouthguards, and the table on the bottom-right lists the mean and the median REBIC of every mouthguard. (A) Brain angle metric (BAM) [14]. (B) Brain injury criteria (BrIC) [27]. (C) Principal component score (PCS) [9].

# Tables:

**Table 1.** Table of abbreviations (in order of appearance).

| Abbreviation | Definition |
|---|---|
| MRE | Mean relative error |
| ATD | Anthropomorphic test dummy |
| mTBI | Mild traumatic brain injury |
| MLSH | Mandible Load Sensing Headform |
| CoG | Center of gravity |
| MiG-C | Stanford's customized mouthguard |
| MiG-B | Stanford's boil-and-bite mouthguard |
| PRE-C | Prevent Biometrics' customized mouthguard |
| PRE-B | Prevent Biometrics' boil-and-bite mouthguard |
| SWA-C | Sports & Wellbeing Analytics' customized mouthguard |
| REPM | Relative error in the peaks of the magnitude |
| CCM | Correlation coefficients of magnitude |
| IAE | Instantaneous axis error |
| CNN | Convolutional neural network |
| FEA | Finite Element Analysis |
| MPS | Maximum principal strain |
| FS | Fiber strain |
| CC | Corpus callosum |
| REBIC | Relative errors in brain injury criteria |
| BAM | Brain angle metric |
| BrIC | Brain injury criteria |
| PCS | Principal component score |

**Table 2.** Parameters and specifications corresponding to the MiG, PRE, and SWA mouthguards, as well as the reference ATD. In time windows after the alignment processing, t=0 ms corresponds to the trigger-point of the ATD (absolute value of linear acceleration at CoG in any components reaches 10g), see Material and Methods section for information about the time alignment.

|  | Stanford mouthguards (MiG-C and MiG-B) | Prevent mouthguards (PRE-C and PRE-B) | Sports & Wellbeing Analytics mouthguard (SWA-C) | ATD (Reference) |
|---|---|---|---|---|
| **Sampling rate (Accelerometer)** | 1,000 Hz | 3,200 Hz | 1,000 Hz | 100,000 Hz |
| **Sampling rate (Gyroscope)** | 8,000 Hz | 3,200 Hz | 952 Hz | 100,000 Hz |
| **Measurement range (Accelerometer)** | ±400 g | ±200 g | ±200 g | ±500 g |
| **Measurement range (Gyroscope)** | ±70 rad/sec | ±35 rad/s | ±35 rad/s | ±140 rad/s |
| **Output time windows** | [-49,150] ms | [0, 50] ms | [1, 103] ms | [-200, 800] ms |
| **Output coordinate axes direction** | X-front, Y-left, z-top | X-front, Y-right, z-bottom | Not parallel to standard coordinate | X-front, Y-right, z-bottom |
| **Output coordinate origin** | Sensor | Center of Gravity (CoG) | Sensor | Center of Gravity (CoG) |
| **Time windows after alignment processing** | [-48, 151] ms | [-10,40] ms | [-1,101] ms | [-200, 800] ms |

**Table 3.** p-value given by the comparison among different impact directions (Kruskal-Wallis 1-way ANOVA). Relative error of peak magnitude (REPM, **Fig. S5**), correlation coefficient of magnitude (CCM, **Fig. S6**), instantaneous axis error (IAE, **Fig. S7**), relative error of brain strain (REBS, **Fig. S8**), and relative error of brain injury criteria (REBIC, **Fig. S9**) are compared. REBS of the PRE mouthguards are not shown due to time windows being too short for the CNN model [31].

|  |  | 0.05<p | 0.01<p<0.05 | 0.001<p<0.01 | p<0.001 |  |
|---|---|---|---|---|---|---|
|  |  | MiG-C | MiG-B | PRE-C | PRE-B | SWA-C |
| REPM | Ang. Acc. | 0.049 | 0.000 | 0.127 | 0.001 | 0.469 |
|  | Ang. Vel. | 0.000 | 0.000 | 0.002 | 0.000 | 0.000 |
|  | Lin. Acc. CoG | 0.000 | 0.000 | 0.026 | 0.000 | 0.000 |
| CCM | Ang. Acc. | 0.000 | 0.000 | 0.000 | 0.000 | 0.000 |
|  | Ang. Vel. | 0.000 | 0.562 | 0.542 | 0.001 | 0.018 |
|  | Lin. Acc. CoG | 0.000 | 0.000 | 0.001 | 0.000 | 0.000 |
| IAE | Ang. Acc. | 0.162 | 0.105 | 0.000 | 0.000 | 0.042 |
|  | Ang. Vel. | 0.000 | 0.000 | 0.000 | 0.000 | 0.025 |
|  | Lin. Acc. CoG | 0.000 | 0.474 | 0.000 | 0.007 | 0.000 |
| REBS | 95% MPS | 0.000 | 0.034 | - | - | 0.687 |
|  | 95% MPS CC | 0.000 | 0.000 | - | - | 0.004 |
|  | 95% FS CC | 0.004 | 0.000 | - | - | 0.001 |
| REBIC | BAM | 0.066 | 0.000 | 0.127 | 0.005 | 0.410 |
|  | BrIC | 0.000 | 0.000 | 0.020 | 0.000 | 0.023 |
|  | PCS | 0.023 | 0.000 | 0.014 | 0.002 | 0.004 |

**Table 4.** p-value given by the comparison among different impact velocities (Kruskal-Wallis 1-way ANOVA). Relative error of peak magnitude (REPM, **Fig. S10**), correlation coefficient of magnitude (CCM, **Fig. S11**), instantaneous axis error (IAE, **Fig. S12**), relative error of brain strain (REBS, **Fig. S13**) and relative error of brain injury criteria (REBIC, **Fig. S14**) are compared. The facemask impacts were not included in this analysis because the two higher impact velocities are not performed as discussed in the text. REBS of PRE mouthguards are not given due to the time windows being too short for the CNN model [31].

| | | MiG-C | MiG-B | PRE-C | PRE-B | SWA-C |
|---|---|---|---|---|---|---|
| REPM | Ang. Acc. | 0.425 | 0.267 | 0.939 | 0.276 | 0.582 |
| | Ang. Vel. | 0.621 | 0.823 | 0.000 | 0.934 | 0.280 |
| | Lin. Acc. CoG | 0.716 | 0.566 | 0.327 | 0.111 | 0.109 |
| CCM | Ang. Acc. | 0.102 | 0.111 | 0.146 | 0.571 | 0.547 |
| | Ang. Vel. | 0.958 | 0.132 | 0.000 | 0.000 | 0.069 |
| | Lin. Acc. CoG | 0.006 | 0.452 | 0.133 | 0.000 | 0.166 |
| IAE | Ang. Acc. | 0.635 | 0.265 | 0.952 | 0.453 | 0.160 |
| | Ang. Vel. | 0.856 | 0.612 | 0.125 | 0.371 | 0.000 |
| | Lin. Acc. CoG | 0.332 | 0.593 | 0.128 | 0.001 | 0.539 |
| REBS | 95% MPS | 0.267 | 0.024 | - | - | 0.944 |
| | 95% MPS CC | 0.077 | 0.007 | - | - | 0.826 |
| | 95% FS CC | 0.043 | 0.233 | - | - | 0.002 |
| REBIC | BAM | 0.520 | 0.751 | 0.771 | 0.551 | 0.299 |
| | BrIC | 0.994 | 0.771 | 0.021 | 0.384 | 0.012 |
| | PCS | 0.002 | 0.181 | 0.669 | 0.063 | 0.067 |

Color legend: 0.05<p | 0.01<p<0.05 | 0.001<p<0.01 | p<0.001

# Supporting Information for Validation and Comparison of Instrumented Mouthguards for Measuring Head Kinematics and Assessing Brain Deformation in Football Impacts

Yuzhe Liu[1,*], August G. Domel[1,*], Seyed Abdolmajid Yousefsani[1,*], Jovana Kondic[1,2]
Gerald Grant[3,4], Michael Zeineh[5], David B. Camarillo[1,3,6,†]

## S1. Instrumented Mouthguard Information

Stanford mouthguards, MiG-B and MiG-C, have the same printed circuit board (PCB), which uses a triaxial accelerometer (H3LIS331DL, ST Microelectronics, Geneva, Switzerland) and a triaxial gyroscope (ITG-3701, InvenSense Inc., San Jose, CA, US). Both of the sensors rest in front of the incisors and are roughly aligned with the middle of the incisors. The data collection is triggered by a linear acceleration of 10 g in any axis of the accelerometer. The raw data is filtered at 160 Hz (4$^{th}$-order Butterworth filter), and then the angular velocity is downsampled to the same time sequence as the linear acceleration. A 5-point stencil derivative of the angular velocity is obtained to calculate the angular acceleration.

The Prevent Biometrics mouthguards, PRE-B and PRE-C, have the same PCB including a triaxial accelerometer (ADXL372, Analog Devices, Boston MA) and a gyroscope (BMG250, Bosch, Gerlingen Germany), and the sensors rest near the first left lateral incisor. The data collection is triggered by a linear acceleration of 5 g in any axis of the accelerometer, and the data is filtered at 200 Hz (4$^{th}$-order Butterworth filter).

The SWA-C mouthguard uses a tri-axial accelerometer (H3LIS331DL, STMicroelectronics, Genova, Switzerland) and a tri-axial gyroscope (LSM9DS1, STMicroelectronics, Genova, Switzerland), and the sensors rest near the left molar. The data collection is triggered by a linear acceleration of 10 g in any direction of the accelerometer. No filters were applied to the raw data and the angular acceleration is derived from the angular velocity using a 5-point stencil derivative. It should be noted that the SWA-C mouthguard was initially developed for rugby.

## S2. Data Processing and Analysis
### S2.1. Reference Data

The raw outputs of the instrumented mouthguards and the reference ATD sensors are angular velocities and linear accelerations, and the coordinates of this output data are listed in **Table 2.** Mouthguard raw data was filtered and processed by each individual mouthguard's firmware, and the reference raw data was filtered by 300Hz (4$^{th}$-order Butterworth filter) according to the standard set forth for the Hybrid III ATD [1]. Reference angular acceleration was then calculated by the 5-point stencil derivative on the filtered angular velocity. To allow for direct comparison with the reference data output, all mouthguard outputs were transformed to the standard ATD coordinate system (+X points forward with respect to the ATD headform; +Y points right; +Z points downward [1]) with the origin at the ATD's CoG.

### S2.2. Kinematic Assessment

The ATD sensors are triggered during an impact when any axis of obtained linear acceleration at CoG exceeds +/- 10g, which serves as a signal to record the given impact. Due to a difference in the positioning of the accelerometers of the mouthguards and the ATD, slightly different trigger times can be observed. Additionally, sampling frequency as well as the duration of impact recording varies across the mouthguards, resulting in different sampling time windows of data for all of the mouthguards. Thus, the following steps were performed to align each mouthguard-obtained output with its reference ATD output, in order to allow for meaningful comparison:

1. The ATD data was interpolated to have the same time resolution as the mouthguard data.
2. The pairwise linear correlation of the magnitude of kinematics within the mouthguard sampling time windows were calculated, and mouthguard data was aligned to the ATD data by finding the highest correlation coefficients.
3. The interpolated ATD data was then cut according to the maximum sampling time windows output for each individual mouthguard.

After performing the time alignment, the mouthguard data and the ATD data had an equivalent time resolution, time window, and trigger time point. The relative error in the peaks of the magnitude (REPM) and the correlation coefficients of magnitude (CCM) (obtained in Step 2 above) were ultimately used as two sets of metrics to assess the accuracy of the mouthguard-obtained measurement with respect to the reference head kinematics.

In addition to the magnitudes of the obtained kinematics, the impact direction is another critical factor in understanding the underlying effects of the impact. Thus, the differences in the axes of the angular velocity, angular acceleration, and the direction of the linear acceleration at CoG were also analyzed. Since the kinematics are measured in the sensors' coordinates, we transformed the measurements obtained by the mouthguards and the ATD to the global coordinate frame. This allows the instantaneous axis error (IAE) to be calculated as the angle between the vectors of kinematics given by the mouthguard and the ATD at the same time point [3]. In this study, the IAE at the time point corresponding to the peak of the magnitude given by the ATD is used as an accuracy metric for assessing the recorded directions of the head impacts.

### S2.3. Brain Deformation Assessment

Head kinematics obtained by the instrumented mouthguards are often used as an input to calculate the extent of brain deformation during an impact. Thus, it is crucial to understand how the error among mouthguard-obtained head kinematics measurements translates to the brain deformation analyses [3]. To compare the mouthguard and the ATD measurements for such a large number of impacts, a validated convolutional neural network (CNN) – based head model [6] was used to calculate the relative error in the brain strain (REBS), as opposed to the traditional finite element model of the brain. The CNN-based brain model [6] was previously trained using an on-field dataset and has been validated to provide a similar brain deformation prediction as the Worcester head injury model (WHIM) model (V1.0) [7]. The outputs of the CNN-based brain model are the 95% maximum principal strain (95% MPS), the 95% maximum principal strain at the corpus callosum (95% MPS at CC), and the 95% fiber strain in the corpus callosum (95% FS at CC). The relative errors for these three strains are used as a test metric to assess the accuracy of the mouthguards.

The input to the CNN-based brain model is the angular velocity in a 100 ms time window with a time resolution of 1 ms. As shown in **Table 2**, the MiG-B, MiG-C, and SWA-C mouthguards' data have time resolutions of 1 ms, and their sampling time windows are longer than the CNN-based brain models' requirements; their data was, thus, cut to 100 ms. The time window of the PRE mouthguards is shorter than required and constant angular velocities were, thus, added after the final recorded data point to reach the mandatory 100 ms input. This leads to the CNN-based brain model substantially overestimating the brain strain. As an example, for the PRE-C mouthguard, the REBS for the 95% MPS was 22.7%, 95% MPS at CC was 18.5%, and 95% FS at CC was 31.5%. To evaluate the error introduced by adding on the artificial constant angular velocity, we cut the ATD data in the PRE-C tests to have the same sampling time windows as the PRE-C mouthguard, added on constant angular velocity to the ATD data in the same manner as done with the PRE-C mouthguard, and then processed everything using the CNN-based brain model. Using this method, the brain strain was still substantially overestimated: (95% MPS: 25.3%; 95% MPS at CC: 21.5%; 95% FS at CC: 30.4%). Therefore, the PRE mouthguards are not analyzed with regard to the brain strain due to their short time window.

### S2.4. Brain Injury Criteria Assessment

In order to identify the risk of brain injury based on the measured head impact kinematics, researchers have recently developed several mTBI-related brain injury criteria that use the head kinematics as an input [2, 4, 5]. By comparing the predicted values of these mTBI-related brain injury criteria using both the ATD and the mouthguard data as an input, the relative errors in brain injury criteria (REBIC) were calculated and used as a set of assessment metrics. In particular, three different mTBI criteria were used in this study: (1) the Brain Angle Metric (BAM), a 3 degree-of-freedom lumped-parameter brain model reflecting the natural frequencies of a finite element brain model [4], (2) the National Highway Traffic Safety Administration (NHTSA)-developed Brain Injury Criteria (BrIC), predicting the risk of mTBI by relating the angular velocity to the critical brain strains [5], and (3) the weighted principal component score (PCS), predicting the mTBI risk by combining previous brain injury criteria developed for severe TBI with weighted coefficient calculations based on football data [2]. All three of the criteria were used in this study to compare the values predicted by the mouthguards to the values predicted by the reference ATD.

## S3. Component traces of angular acceleration, angular velocity and linear acceleration (corresponding to Fig.2)

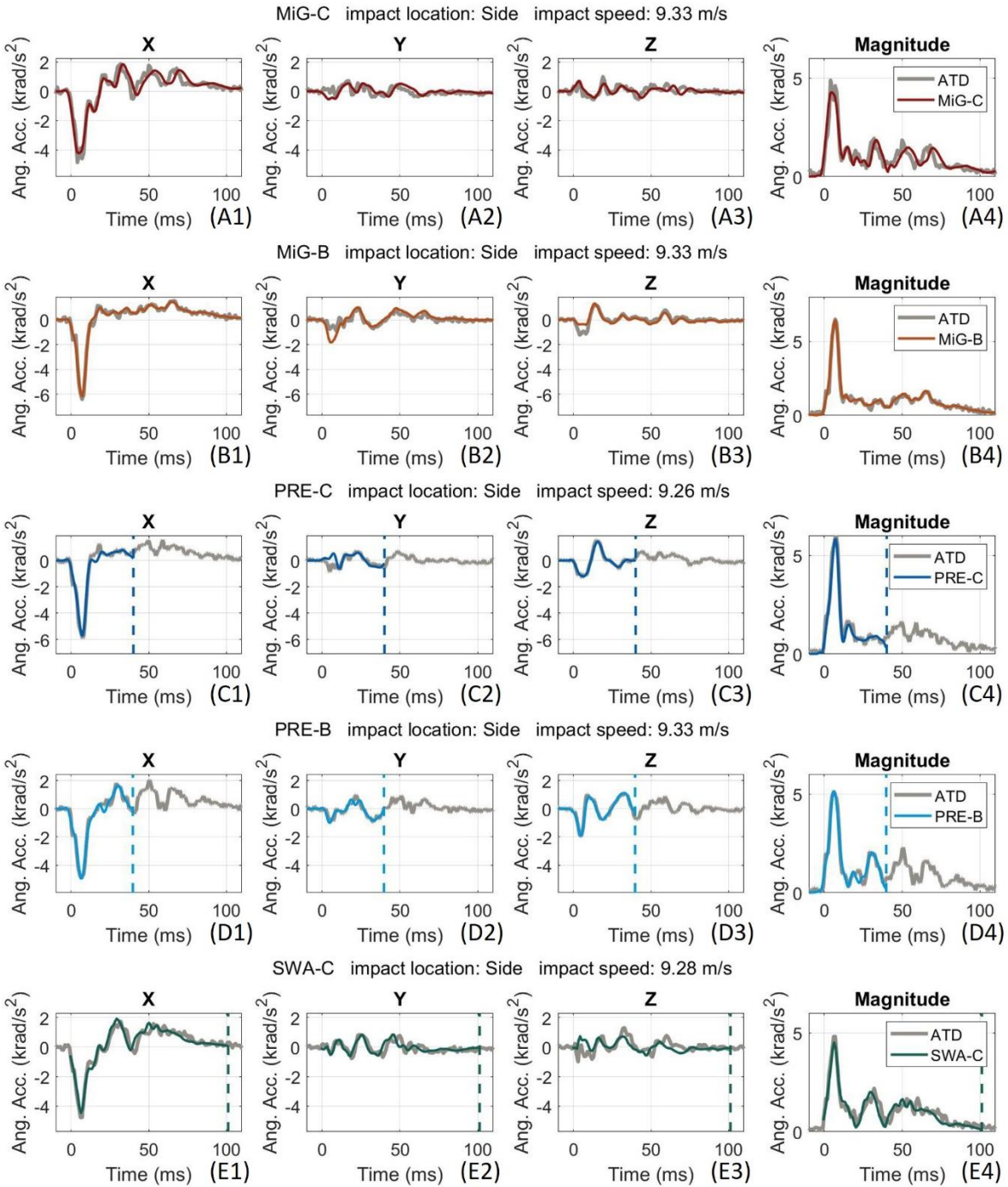

**Figure S1.** Component traces of angular acceleration for ~9.3 m/s side impact. (A) MiG-C, (B) MiG-B, (C) PRE-C, (D) PRE-B, (E) SWA-C; (1) X direction (from back to front), (2) Y direction (from left to right), (3) Z direction (from top to bottom), (4) magnitude.

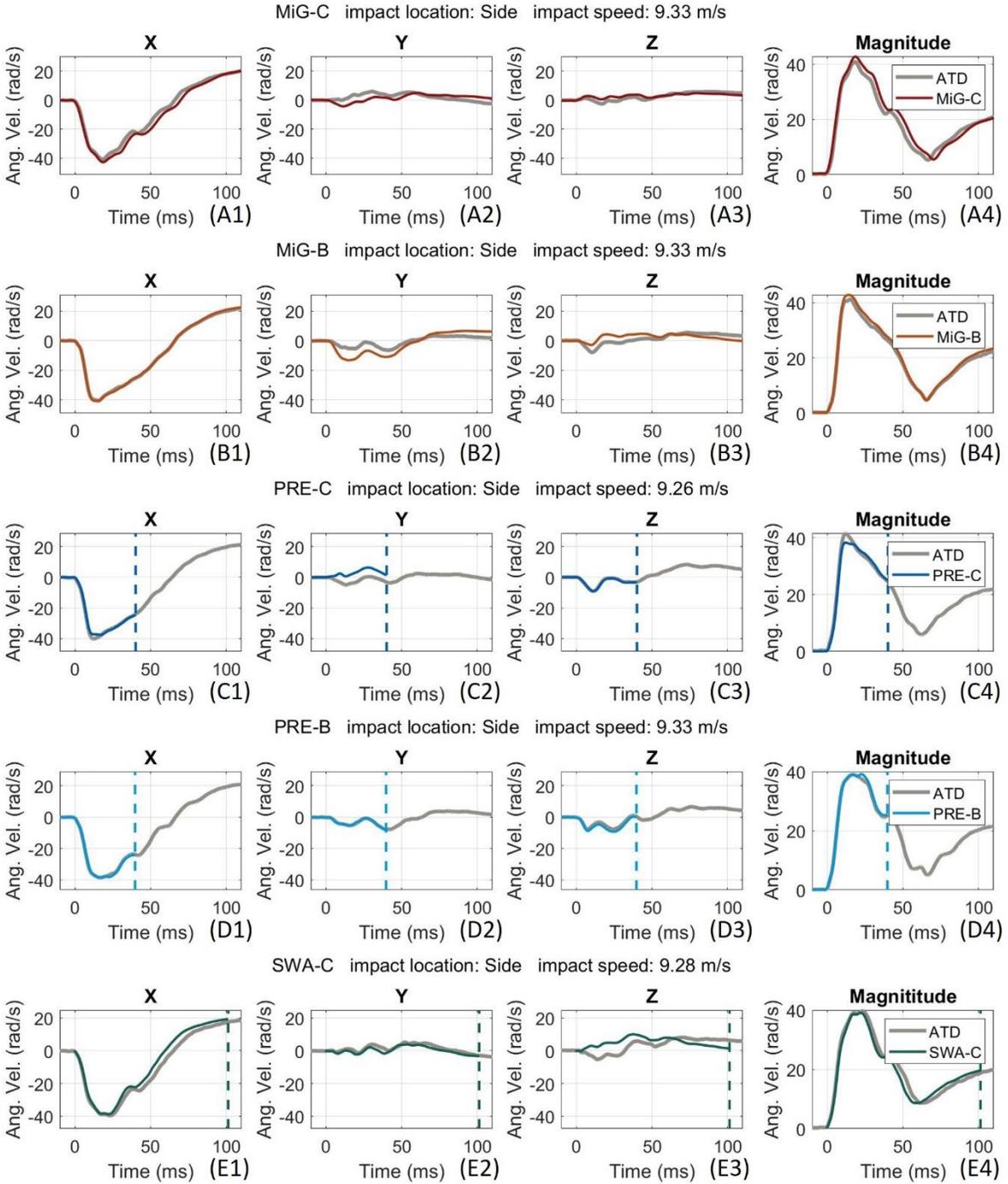

**Figure S2.** Component traces of angular velocity for ~9.3 m/s side impact. (A) MiG-C, (B) MiG-B, (C) PRE-C, (D) PRE-B, (E) SWA-C; (1) X direction (from back to front), (2) Y direction (from left to right), (3) Z direction (from top to bottom), (4) magnitude.

54

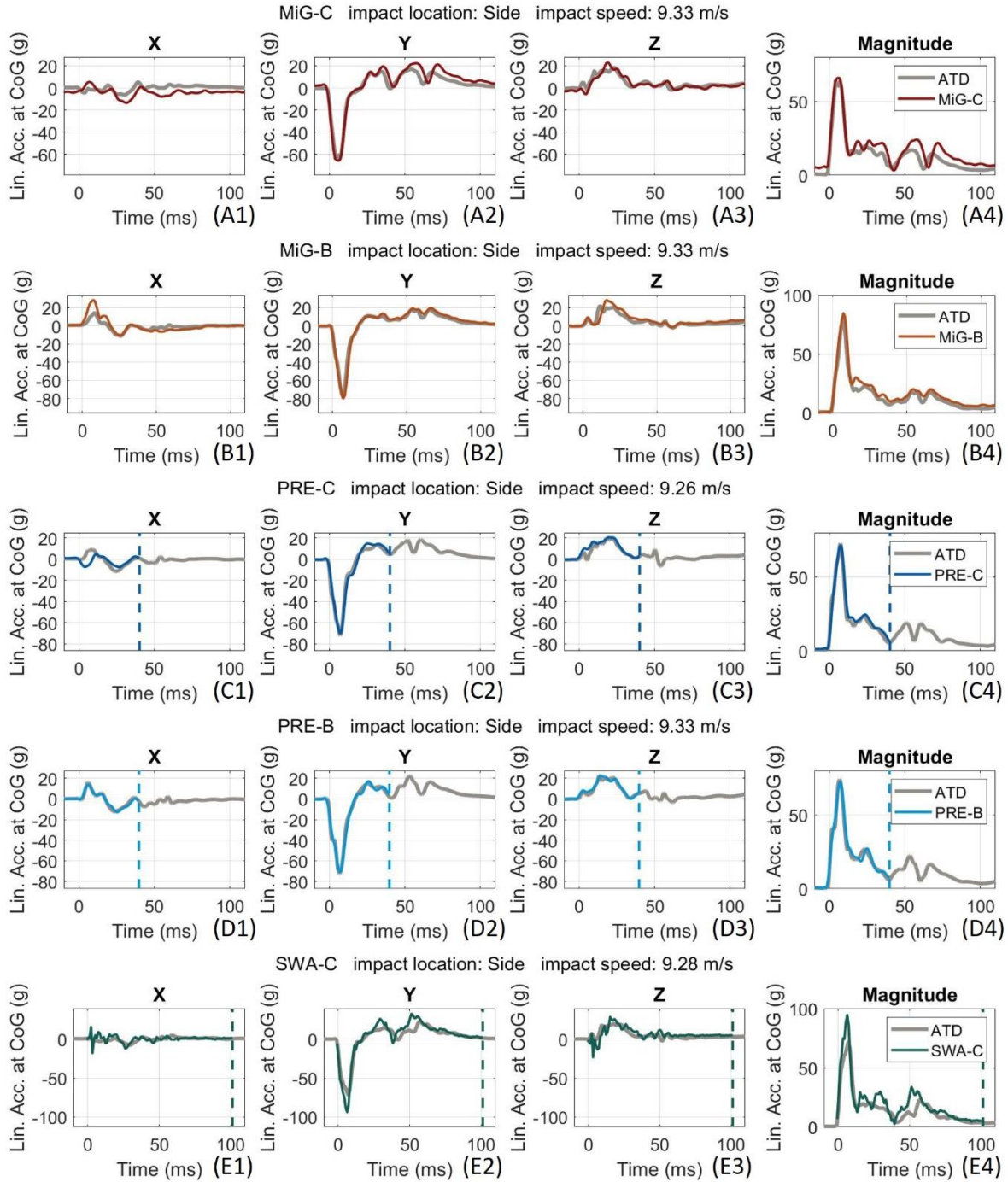

**Figure S3.** Component traces of linear acceleration at CoG for ~9.3 m/s side impact. (A) MiG-C, (B) MiG-B, (C) PRE-C, (D) PRE-B, (E) SWA-C; (1) X direction (from back to front), (2) Y direction (from left to right), (3) Z direction (from top to bottom), (4) magnitude.

55

## S4. Transformation of linear acceleration from the sensor to CoG

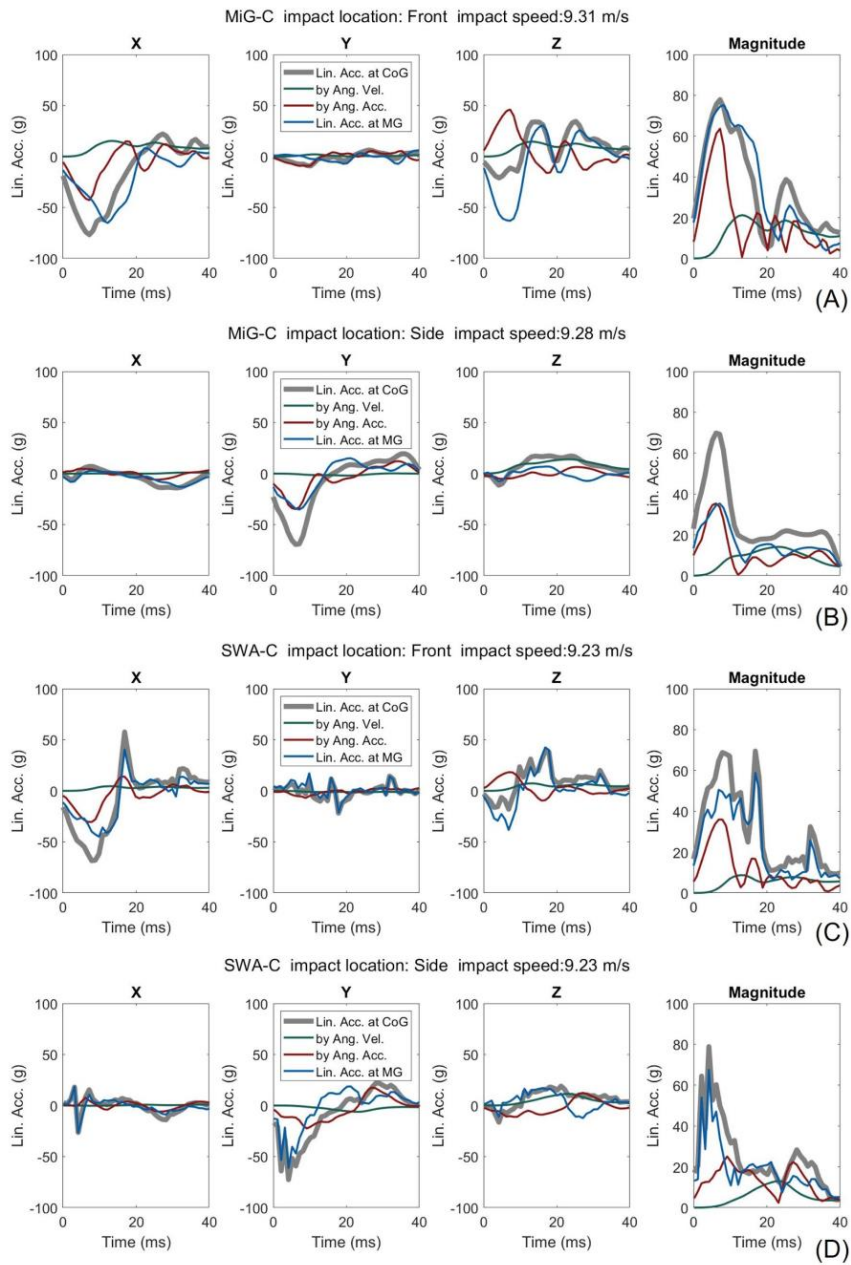

**Figure S4.** Linear acceleration at CoG, linear acceleration at the location corresponding to the mouthguard, linear acceleration translated to CoG contributed by angular acceleration (by Ang. Acc.), linear acceleration translated to CoG contributed by angular velocity (by Ang. Vel.). (A) Front impact in MiG-C test; (B) side impact in MiG-C test; (C) front impact in SWA-C test; (D) side impact in SWA-C test.

## S5. Comparison among impact locations

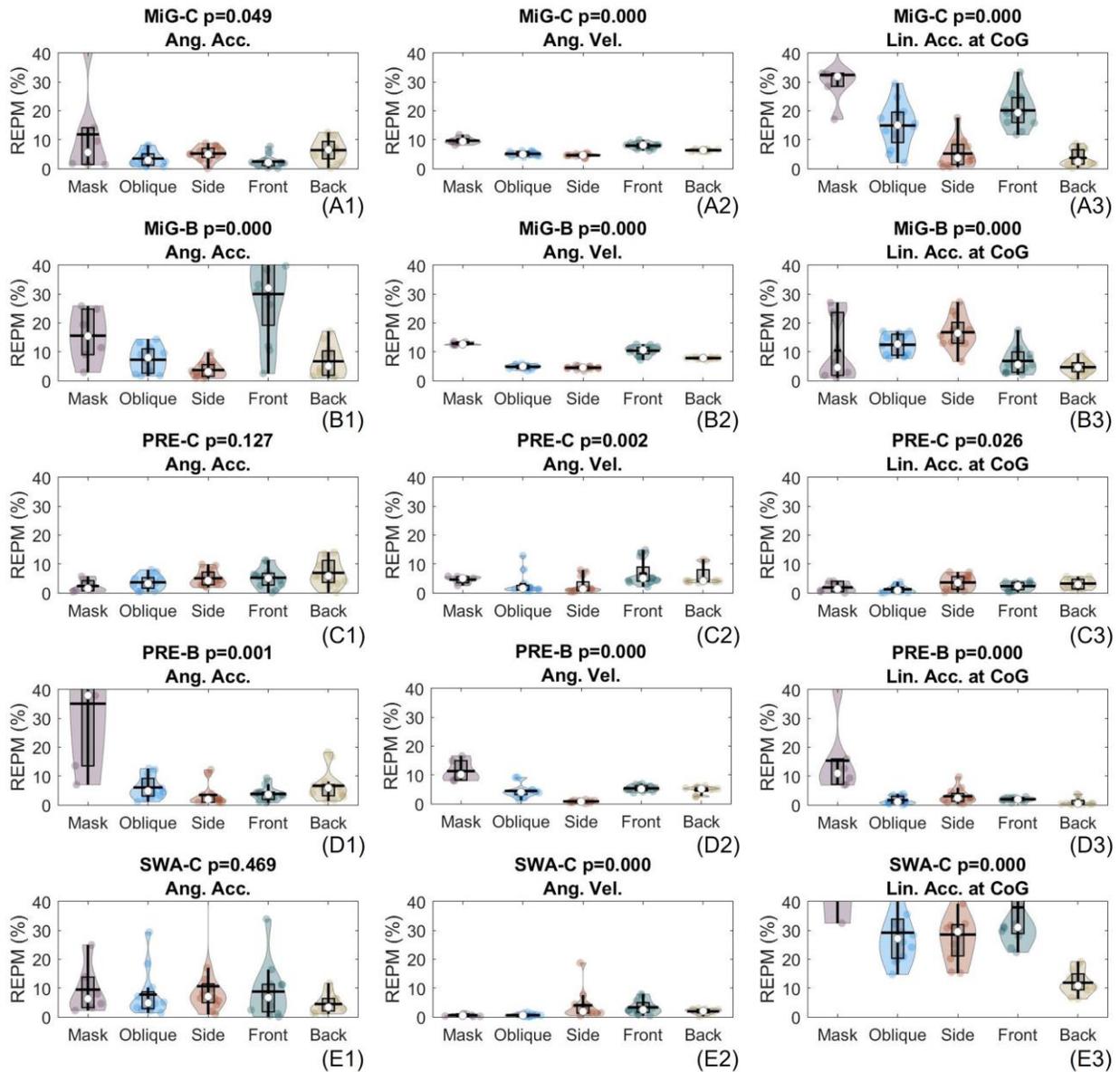

**Figure S5.** The effect of impact location on the relative error in peak of the magnitude (REPM) for each mouthguard.

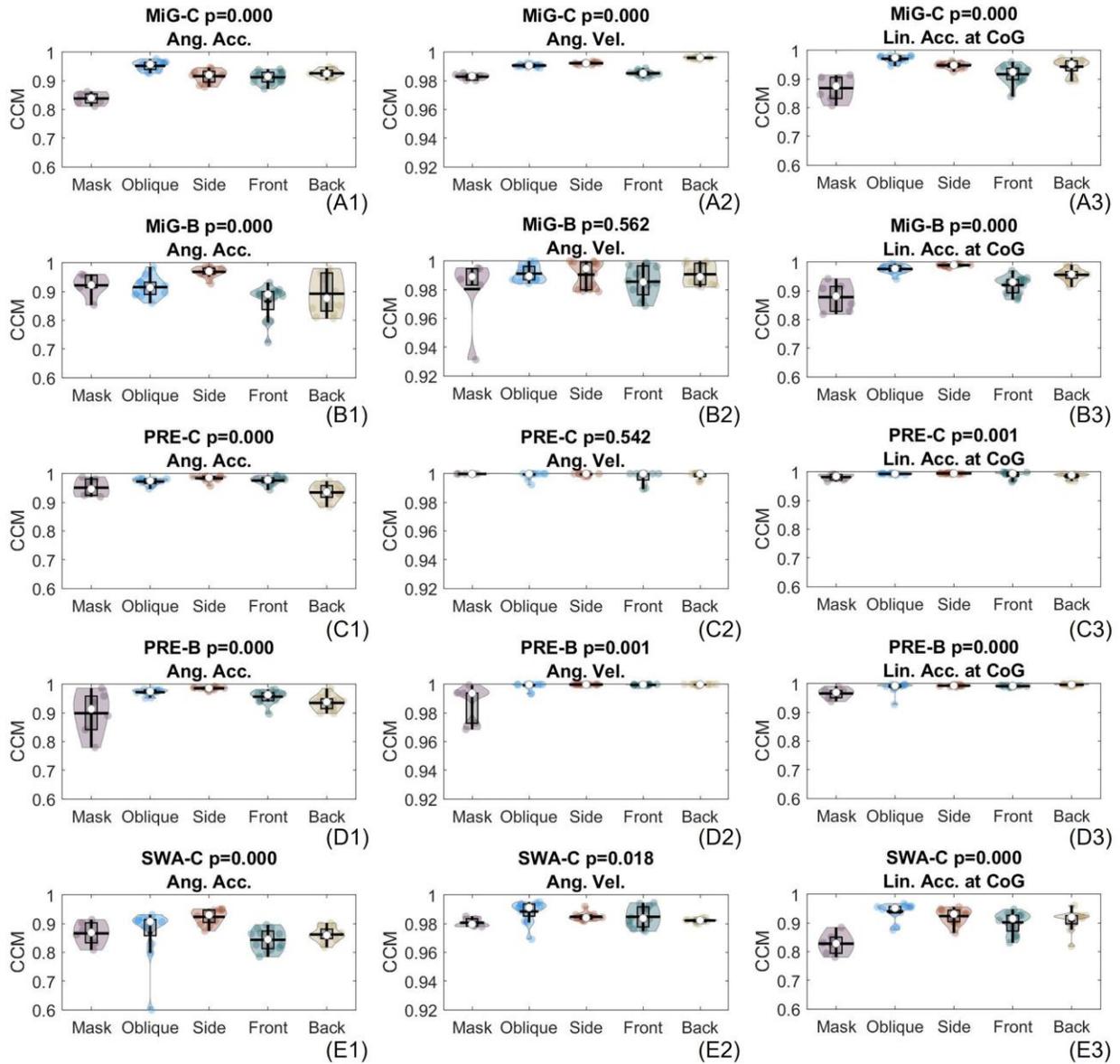

**Figure S6.** The effect of impact location on the correlation coefficient of the magnitude (CCM) for each mouthguard.

58

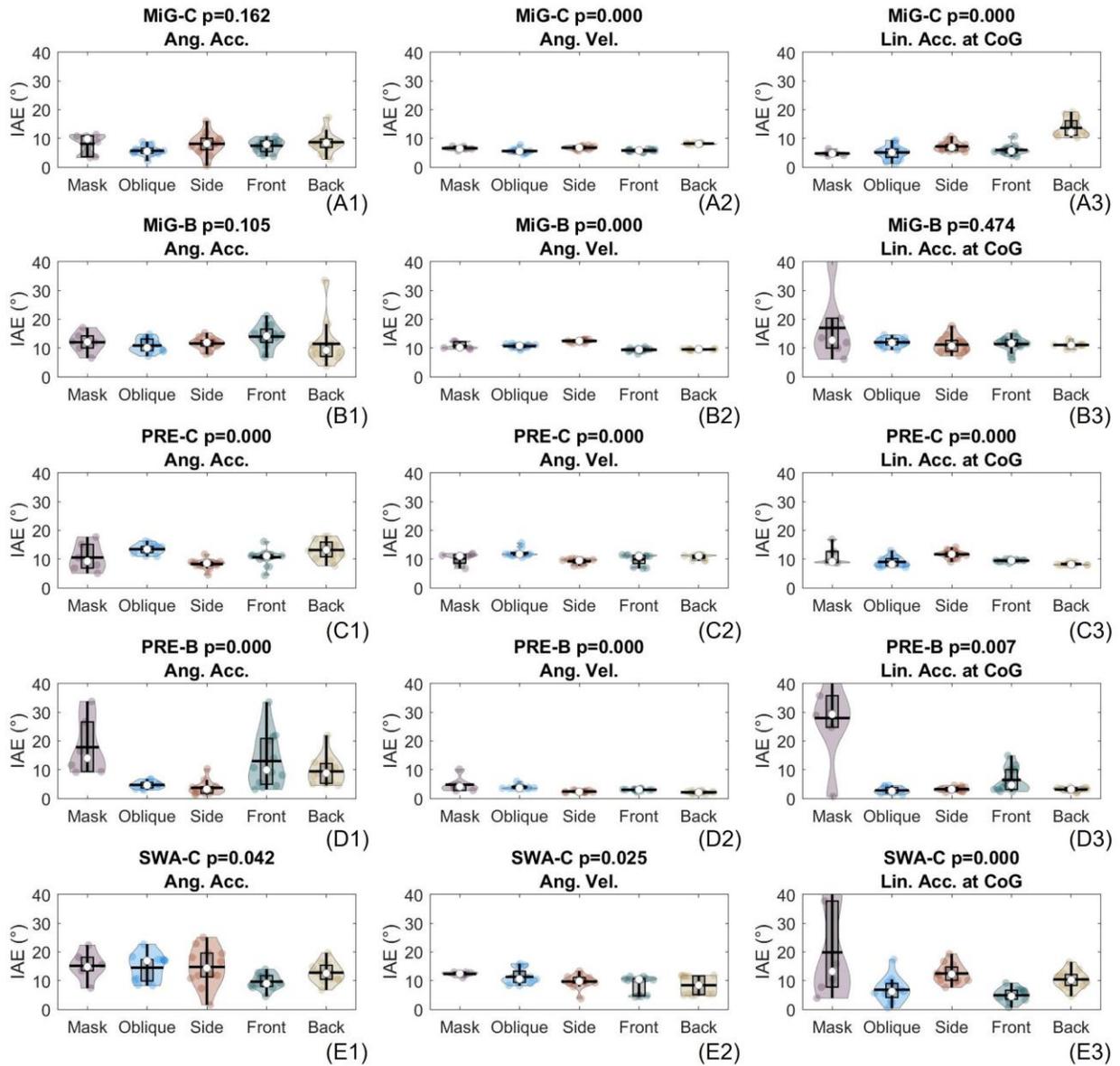

**Figure S7.** The effect of impact location on the instantaneous axis error (IAE) for each mouthguard.

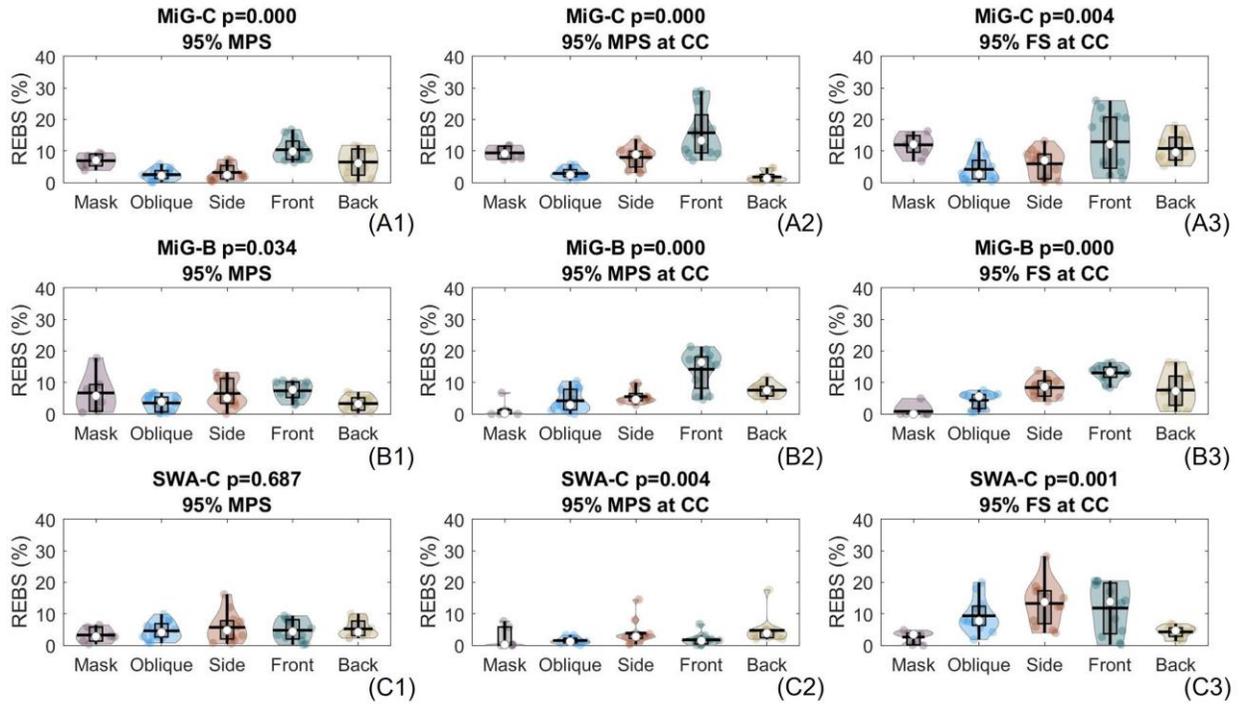

**Figure S8.** The effect of impact location on the relative error in the brain strain (REBS) for each mouthguard.

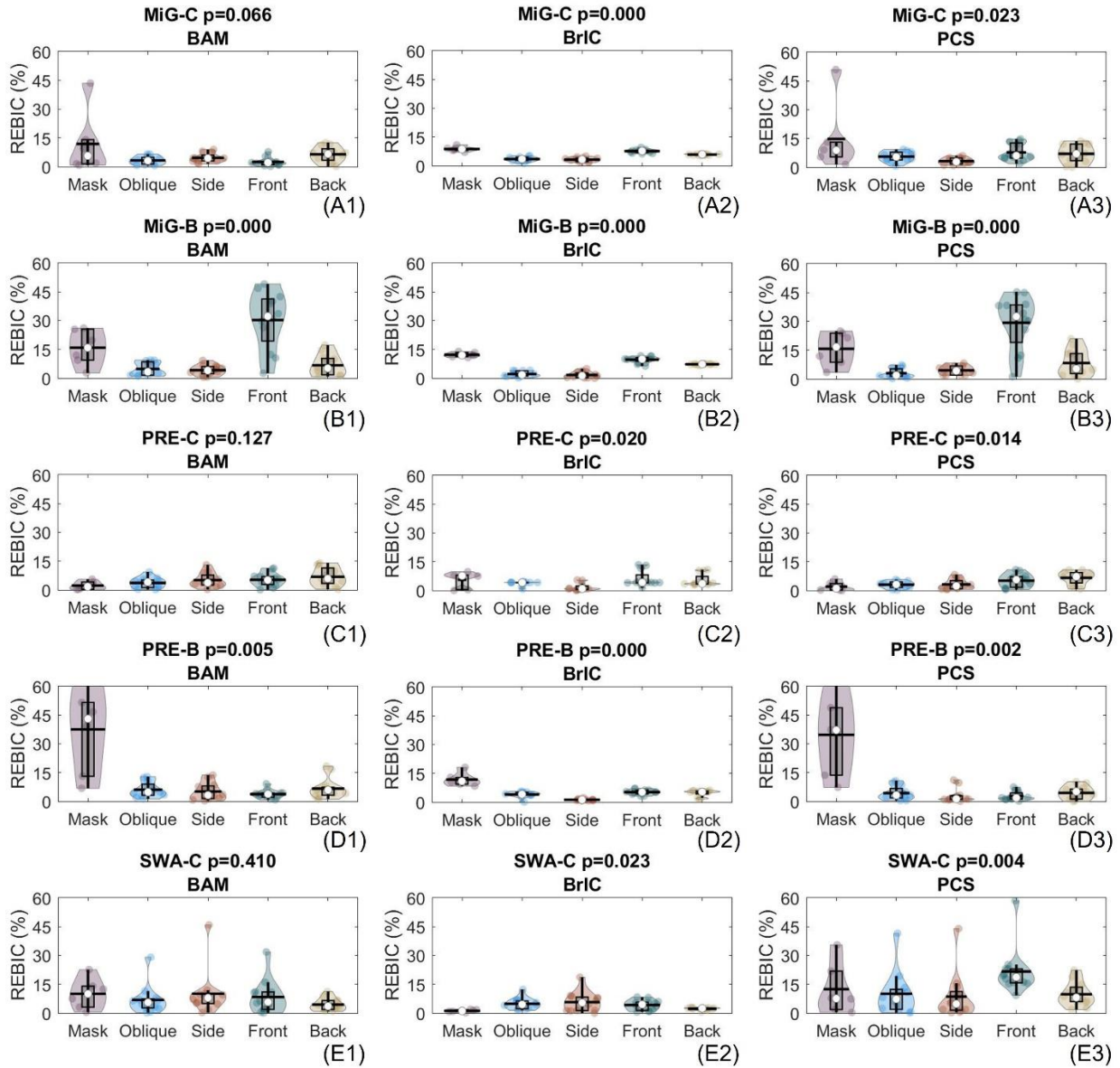

**Figure S9.** The effect of impact location on the relative error in brain injury criterion (REBIC) for each mouthguard.

## S6. Comparison among impact velocities

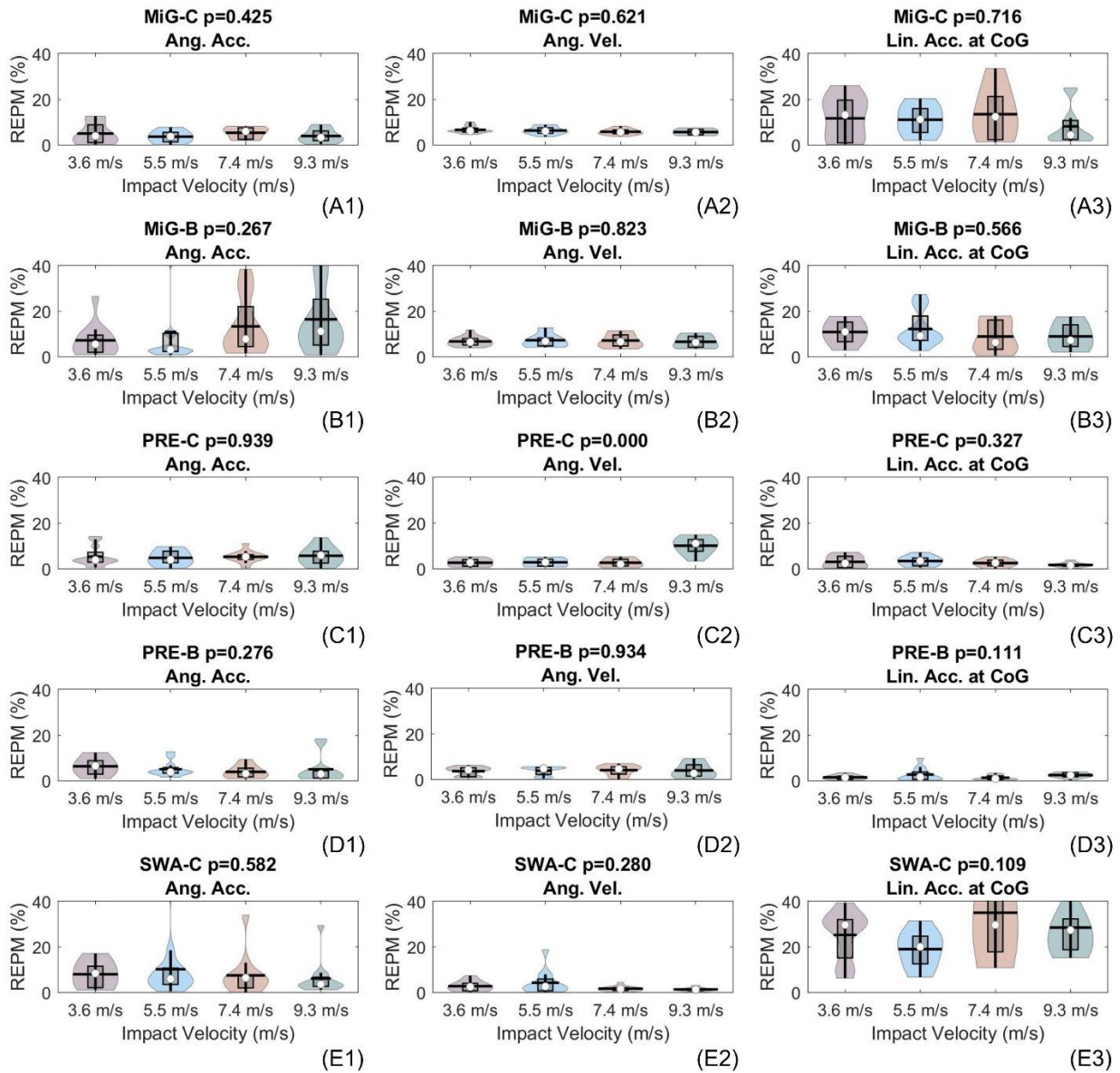

**Figure S10.** The effect of impact velocity on the relative error in peak of the magnitude (REPM) for each mouthguard. Colors of points refer to different impact locations (legend is at the top of the figure).

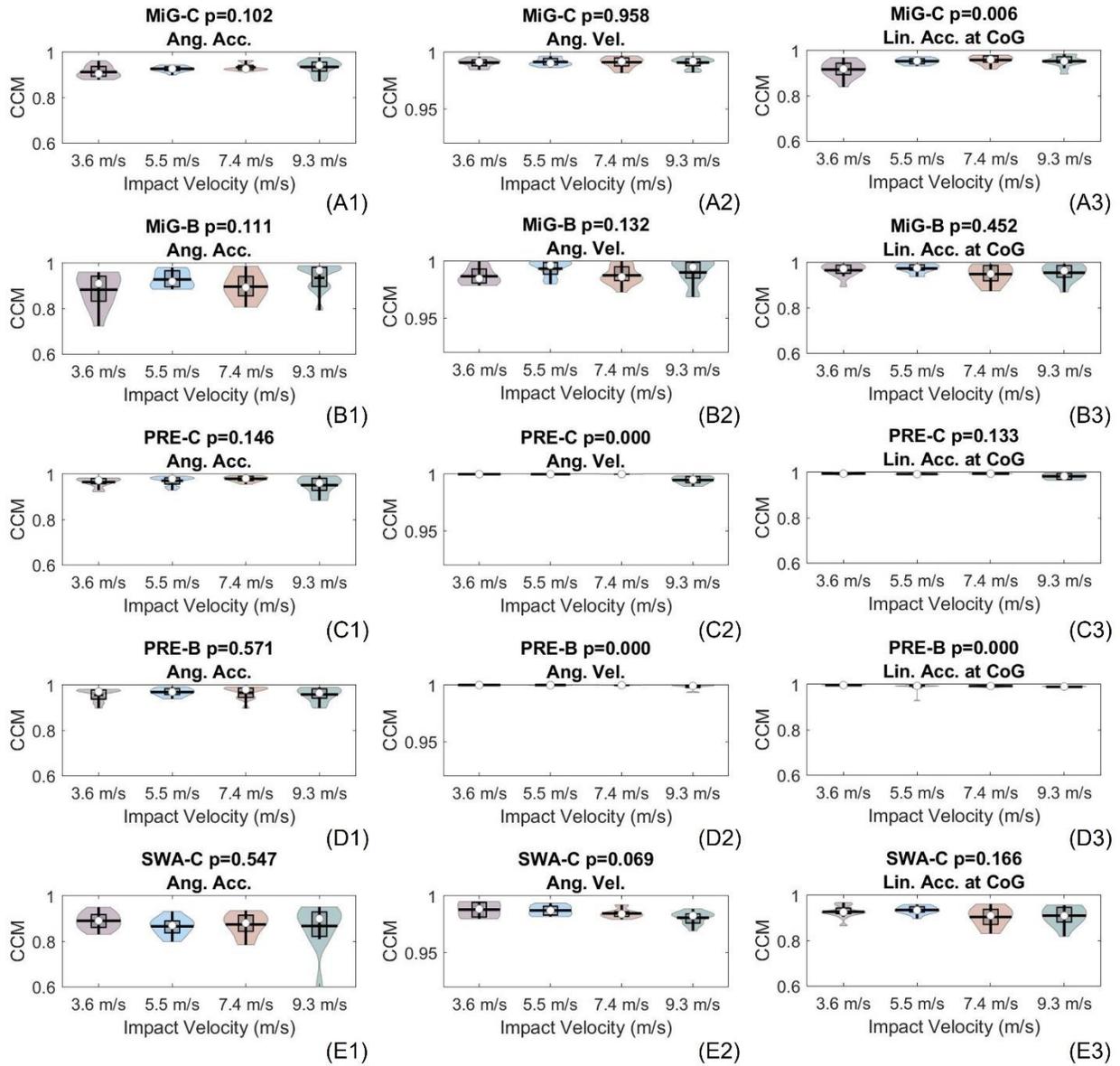

**Figure S11.** The effect of impact velocity on the correlation coefficient of the magnitude (CCM) for each mouthguard. Colors of points refer to different impact locations (the legend is at the top of the figure).

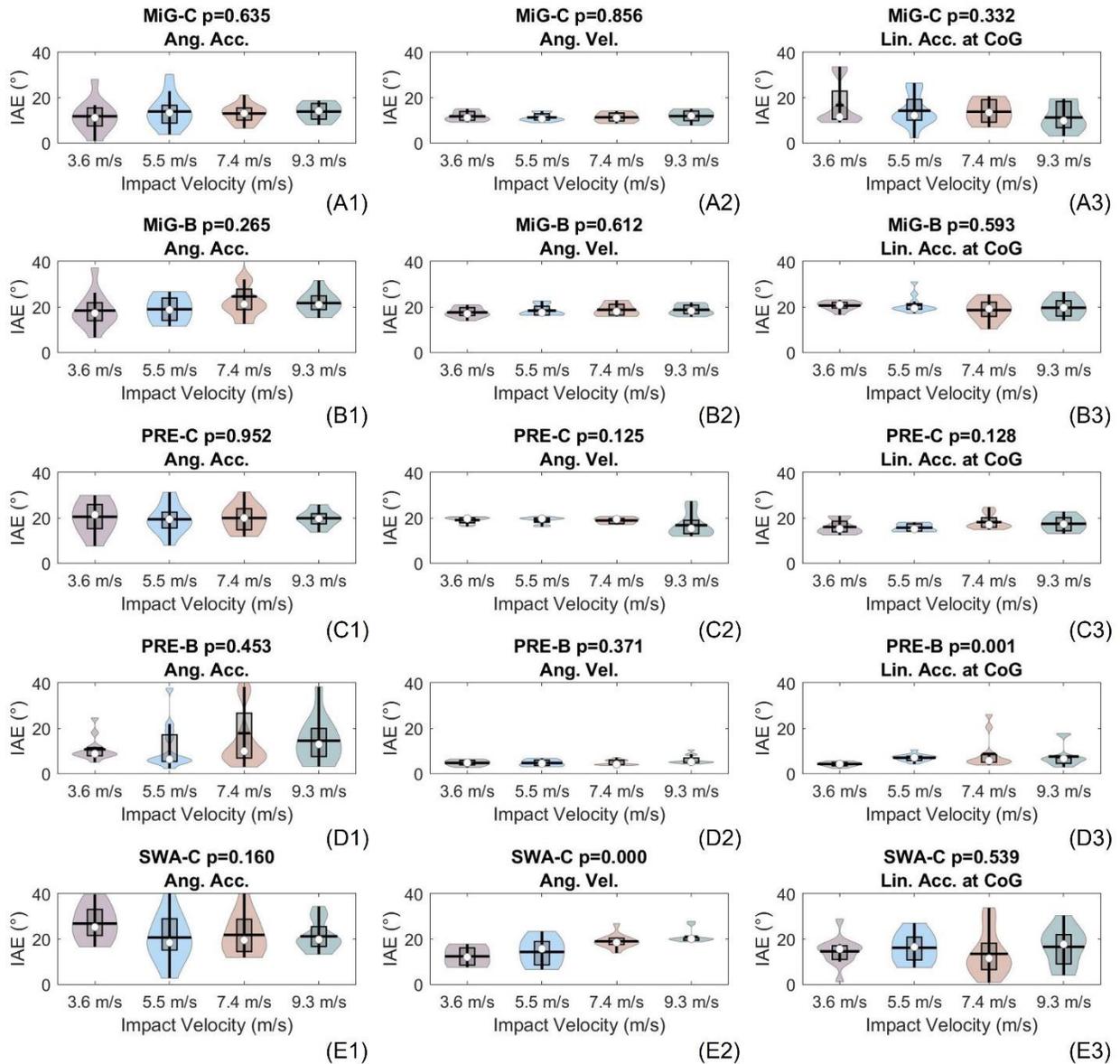

**Figure S12.** The effect of impact velocity on the instantaneous axis error (IAE) for each mouthguard. Colors of points refer to different impact locations (the legend is at the top of the figure).

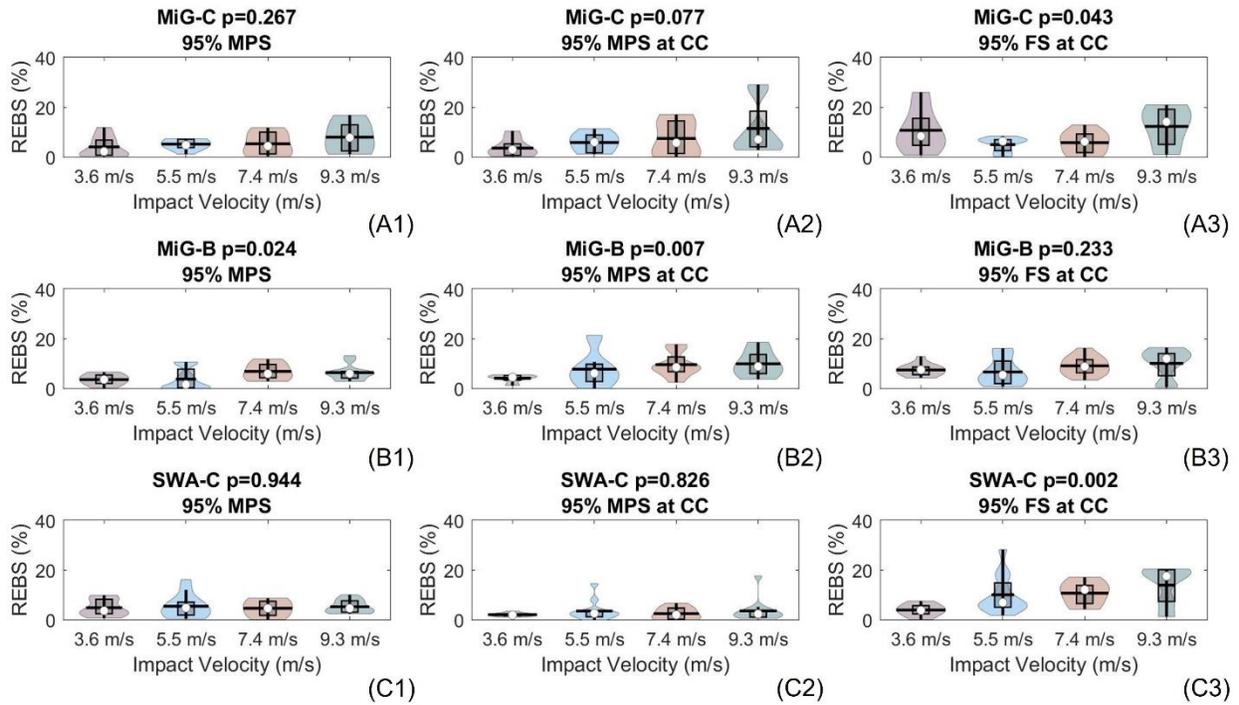

**Figure S13.** The effect of impact velocity on the relative error in the brain strain (REBS) for each mouthguard. Colors of points refer to different impact locations (the legend is at the top of the figure).

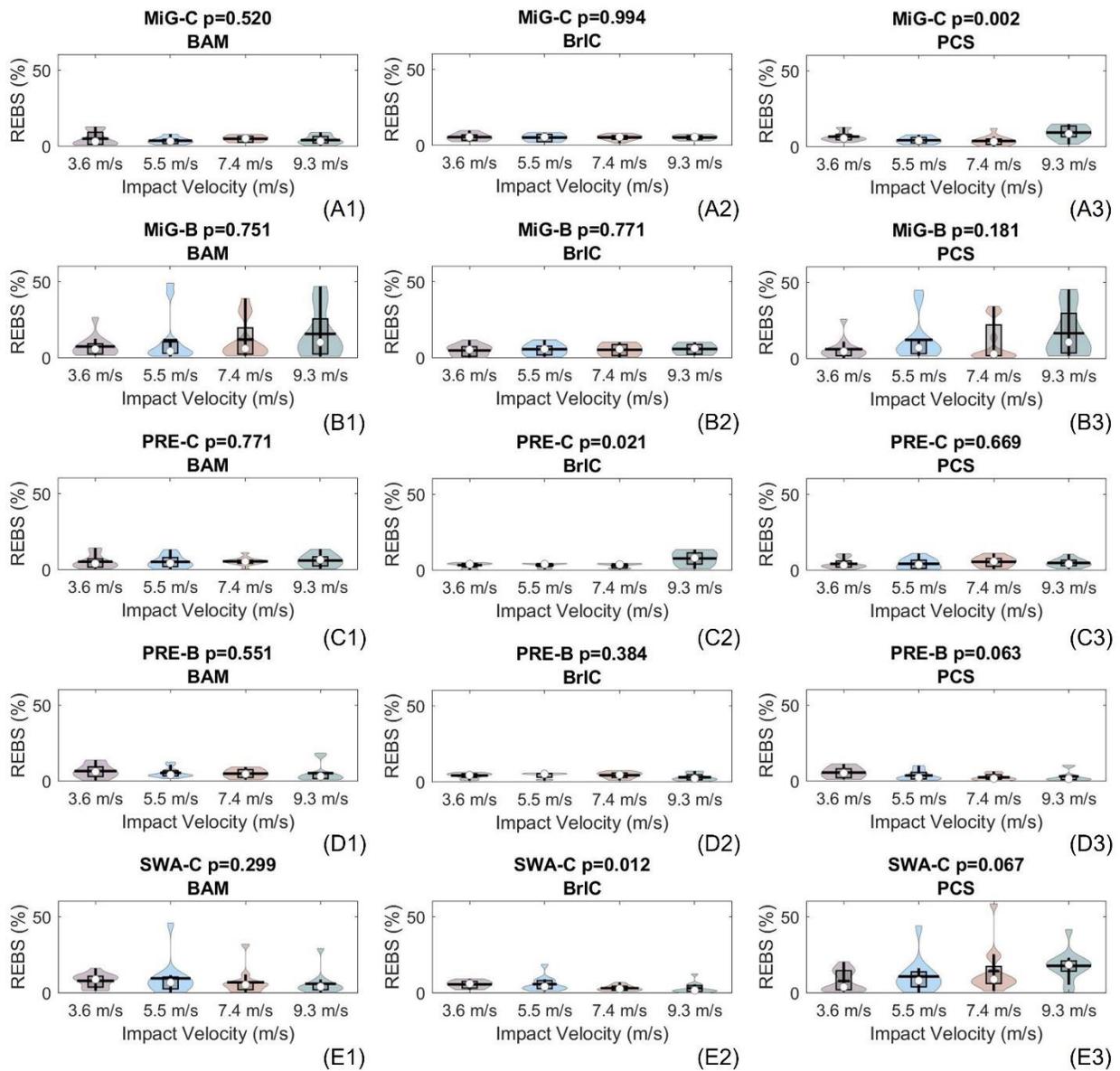

**Figure S14.** The effect of impact velocity on the relative error in brain injury criterion (REBIC) for each mouthguard. Colors of points refer to different impact locations (the legend is at the top of the figure).